\documentclass[12pt]{article}
\usepackage[dvips]{graphicx}

\usepackage{epsfig,amsmath,amssymb,verbatim,mathrsfs}
\def\beq{\begin{eqnarray}}
\def\eeq{\end{eqnarray}}
\def\bea{\begin{eqnarray}}
\def\eea{\end{eqnarray}}

\def\gev{\, {\rm GeV}}
\def\mev{\, {\rm MeV}}

\newcommand{\gsim}{\lower.7ex\hbox{$\;\stackrel{\textstyle>}{\sim}\;$}}
\newcommand{\lsim}{\lower.7ex\hbox{$\;\stackrel{\textstyle<}{\sim}\;$}}

\def\stilde{\widetilde}

\newcommand{\newc}{\newcommand}
\newc{\Nc}{N_{c}}
\newc{\CG}{C_G}
\newc{\gp}{g'}
\newc{\stopi}{\stilde t_i}
\newc{\sboti}{\stilde b_i}
\newc{\staui}{\stilde \tau_i}
\newc{\stopj}{\stilde t_j}
\newc{\sbotj}{\stilde b_j}
\newc{\stauj}{\stilde \tau_j}
\newc{\stopI}{\stilde t_1}
\newc{\stopII}{\stilde t_2}
\newc{\sbotI}{\stilde b_1}
\newc{\sbotII}{\stilde b_2}
\newc{\stauI}{\stilde \tau_1}
\newc{\stauII}{\stilde \tau_2}
\newc{\sstop}{s_{t}}
\newc{\cstop}{c_{t}}
\newc{\ssbot}{s_{b}}
\newc{\csbot}{c_{b}}
\newc{\sstau}{s_{\tau}}
\newc{\cstau}{c_{\tau}}
\newc{\Sstop}{s_{2t}}
\newc{\Cstop}{c_{2t}}
\newc{\Ssbot}{s_{2b}}
\newc{\Csbot}{c_{2b}}
\newc{\Sstau}{s_{2\tau}}
\newc{\Cstau}{c_{2\tau}}
\newc{\salpha}{s_\alpha}
\newc{\calpha}{c_\alpha}
\newc{\Calpha}{c_{2\alpha}}
\newc{\Salpha}{s_{2\alpha}}
\newc{\sbetapm}{s_{\beta_\pm}}
\newc{\cbetapm}{c_{\beta_\pm}}
\newc{\Sbetapm}{s_{2 \beta_\pm}}
\newc{\Cbetapm}{c_{2 \beta_\pm}}
\newc{\sbetaO}{s_{\beta_0}}
\newc{\cbetaO}{c_{\beta_0}}
\newc{\SbetaO}{s_{2 \beta_0}}
\newc{\CbetaO}{c_{2 \beta_0}}
\newc{\vu}{v_u}
\newc{\vd}{v_d}
\newc{\seL}{\stilde e_L}
\newc{\smuL}{\stilde \mu_L}
\newc{\seR}{\stilde e_R}
\newc{\smuR}{\stilde \mu_R}
\newc{\suL}{\stilde u_L}
\newc{\sdL}{\stilde d_L}
\newc{\suR}{\stilde u_R}
\newc{\sdR}{\stilde d_R}
\newc{\scL}{\stilde c_L}
\newc{\ssL}{\stilde s_L}
\newc{\scR}{\stilde c_R}
\newc{\ssR}{\stilde s_R}
\newc{\snue}{\stilde \nu_e}
\newc{\snumu}{\stilde \nu_\mu}
\newc{\snutau}{\stilde \nu_\tau}
\newc{\Gpm}{G^\pm}
\newc{\Hpm}{H^\pm}
\newc{\FFbS}{\overline{FF}S}
\newc{\FFbV}{\overline{FF}V}
\newc{\FSS}{F_{SS}}
\newc{\FSSS}{F_{SSS}}
\newc{\FFFS}{F_{FFS}}
\newc{\FFFbS}{F_{\overline{FF}S}}
\newc{\FSSV}{F_{SSV}}
\newc{\FVS}{F_{VS}}
\newc{\FVVS}{F_{VVS}}
\newc{\FFFV}{F_{FFV}}
\newc{\FFFbV}{F_{\overline{FF}V}}
\newc{\Fgauge}{F_{\rm gauge}}
\newc{\DRbarprime}{$\overline{\rm DR}'$ }
\newc{\DRbar}{$\overline{\rm DR}$ }
\newc{\MSbar}{$\overline{\rm MS}$ }
\newc{\Yu}{{\bf Y}_u}
\newc{\Yd}{{\bf Y}_d}
\newc{\Ye}{{\bf Y}_e}
\newc{\Au}{{\bf a}_u}
\newc{\Ad}{{\bf a}_d}
\newc{\Ae}{{\bf a}_e}
\newc{\bm}{{\bf m}}
\newc{\zhol}{Z^{\rm hol}}

\newc{\rwino}{r_{\tilde W}}
\newc{\rmu}{r_{\tilde H}}
\newc{\ra}{r_A}
\newc{\ccdot}{\!\cdot\!}

\newcommand{\nnmb}{\nonumber}

\newcommand{\lrf}[2]{\left(\frac{#1}{#2}\right)}

\begin{document}

\setlength{\baselineskip}{0.2in}

%\begin{comment}

%@twocolumnfalse\endcsname
%%
%%
\begin{titlepage}
\noindent
%\begin{flushright}
%\end{flushright}
%%
%%
%\vspace{10cm}
\flushright{October 2010}
\vspace{1cm}

\begin{center}
  \begin{Large}
    \begin{bf}
Low-Energy Signals from Kinetic Mixing with a Warped Abelian Hidden Sector\\
     \end{bf}
  \end{Large}
\end{center}
\vspace{0.2cm}

\begin{center}

\begin{large}
Kristian L. McDonald$^{(a),(b)}$ and David E. Morrissey$^{(a)}$\\
\end{large}
\vspace{1cm}
  \begin{it}
$(a)$ TRIUMF,
4004 Wesbrook Mall, Vancouver, BC V6T 2A3, Canada.\\ \vspace{0.5cm}
$(b)$ Max-Planck-Institut f\"ur Kernphysik,\\
 Postfach 10 39 80, 69029 Heidelberg, Germany.\\
\vspace{0.5cm}
Email: kristian.mcdonald@mpi-hd.mpg.de, dmorri@triumf.ca
\vspace{0.5cm}
\end{it}

\end{center}

%\center{\today}

\begin{abstract}

  We investigate the detailed phenomenology of a light Abelian hidden 
sector in the Randall-Sundrum framework.
Relative to other works with light hidden sectors, 
the main new feature is a tower of hidden Kaluza-Klein 
vectors that kinetically mix with the 
Standard Model photon and $Z$.
We investigate the decay properties of the hidden sector fields
in some detail, 
and develop an approach for calculating processes initiated 
on the ultraviolet brane of a warped space
with large injection momentum relative to
the infrared scale.  
Using these results,
we determine the detailed bounds on the light
warped hidden sector 
from precision electroweak measurements and low-energy experiments.
We find viable regions of parameter space that lead to significant
production rates for several of the hidden Kaluza-Klein vectors in meson 
factories and fixed-target experiments.  
This offers the possibility of exploring the structure of an extra spacetime 
dimension with lower-energy probes.

 \end{abstract}

\vspace{1cm}

\end{titlepage}

% %\setcounter{footnote}{1}
% \setcounter{page}{2}
% %\setcounter{figure}{0}
% %\setcounter{table}{0}

% %\tableofcontents

% \vfill\eject

% %\end{comment}

%%%%%%%%%%%%%%%%%%%%%%%%%%%%%%%%%%%%%%%%%%%%%%%%%%%%%%%%%%%%%%%%%%%%%%

\section{Introduction\label{sec:intro}}

  An interesting possibility for new physics beyond the 
Standard Model~(SM) is a \emph{light hidden sector}, consisting of exotic 
particles with masses well below the electroweak scale and very 
weak couplings to the SM~\cite{Strassler:2006im,Patt:2006fw,Zurek:2010xf}.  
Even though such states could have been created in a number 
of previous and current experiments, their rate of production can be 
consistent with experimental bounds provided their couplings to the SM 
are sufficiently small.  
Light hidden sectors can give rise to new
and unusual signatures, and their traces might already
be present in existing experimental data sets or discoverable in
planned upcoming
searches~\cite{Borodatchenkova:2005ct,Batell:2009yf,Essig:2009nc,Reece:2009un,Bjorken:2009mm}. 
It is therefore important to understand 
the signals of viable low-energy extensions of the SM to ensure that 
maximal use is made of both existing and forthcoming data sets. 

  New physics below the electroweak scale arises in a number
of scenarios extending the SM, and has been proposed as a central
component of several theories of dark matter~\cite{Pospelov:2007mp,
ArkaniHamed:2008qn,Hooper:2008im}. The presence of a 
sub-electroweak scale introduces another separation of scales beyond 
the usual weak/Planck hierarchy and one expects the hidden sector 
to contain some suitable mechanism to ensure radiative stability. 
The standard solutions to the hierarchy problem can be 
considered in this context and the sub-electroweak scale of the hidden
sector could arise from supersymmetry~\cite{ArkaniHamed:2008qn}, 
strong dynamics~\cite{Alves:2009nf}, 
or a warped extra dimension~\cite{Gherghetta:2010cq,McDonald:2010iq}.
The details of the stabilization mechanism can significantly modify 
the resulting experimental signals since they  
can lead to very different particle spectra below the TeV scale. 

  In Ref.~\cite{McDonald:2010iq} we investigated a simple hidden sector with
an Abelian $U(1)_x$ hidden gauge symmetry in an extended Randall-Sundrum 
model~\cite{Randall:1999ee,Randall:1999vf}.  Relative to many other 
realizations of light Abelian hidden sectors,
the model predicted an entire Kaluza-Klein (KK) tower 
of light vector bosons. These modes can have important
phenomenological consequences since several modes in the tower 
can couple significantly to the SM through gauge kinetic mixing.
This feature could allow one to study the structure of the 
warped extra dimension in lower-energy experiments, such as meson factories 
or fixed-target experiments.

  In order to focus specifically on the interesting 
low-energy physics, in the present work we study a slightly simpler 
theory than was presented
in Ref.~\cite{McDonald:2010iq}. We consider a single warped
 bulk containing a $U(1)_x$ gauge theory with the SM confined to the
 ultraviolet~(UV) brane. 
As before, we couple the SM and the hidden sector through a gauge 
kinetic mixing operator localized on the UV brane. 
With an infrared (IR) brane scale of order a GeV this setup reproduces
the interesting low-energy phenomenology of Ref.~\cite{McDonald:2010iq}.
Beyond the TeV scale one should consider the full structure of
Ref.~\cite{McDonald:2010iq} or include some other mechanism to 
stabilize the electroweak scale, but these details are not important as 
far as the 
relevant
low-energy phenomenology is concerned. 
Let us also point out that,
via the AdS/CFT correspondence, this model  
can be considered as a dual description of 
a purely 4D theory 
containing
fundamental SM fields coupled to a hidden conformal field theory~(CFT) 
with a weakly gauged $U(1)_x$ subgroup and an order GeV mass
gap~\cite{ArkaniHamed:2000ds}.

  In the present work we perform a detailed investigation of the 
phenomenology of a simple light
warped hidden sector
%,
with a characteristic mass scale of roughly 10 MeV to 10 GeV. 
This range is technically natural, but much smaller and much larger
hidden mass scales are also possible (and natural).  
We concentrate on these particular values primarily because they
are phenomenologically interesting.\footnote{We note that gauge kinetic 
mixing with a light hidden sector may also be motivated by recent models 
of dark matter~\cite{Pospelov:2007mp,ArkaniHamed:2008qn}. We do not 
consider this matter here, but the phenomenology we consider would 
likely form a central component of warped realizations of these models.} 
In particular, we will show 
that a warped hidden sector in this mass range can be consistent with
existing bounds from direct searches and astrophysics while also giving
rise to potentially observable new signals in lower-energy collider 
experiments.  It is these signals that 
we investigate in the present work.

  The outline of this paper is as follows.  In Section~\ref{sec:rstheory} 
we describe the model in more detail.  Next we discuss the decays
of hidden sector KK modes in Section~\ref{sec:decay}.  
In Section~\ref{sec:match} we address the important issue of reliably
computing processes initiated by the SM
on the UV brane with energies well above the hidden IR scale but
well below the characteristic UV scale. Specifically we discuss the
matching of the 4D KK effective theory 
to the full 5D bulk theory.
We apply these results in Sections~\ref{sec:pew} and \ref{sec:lowe} 
where we compute, respectively, the precision electroweak constraints,
and the bounds and discovery prospects at fixed-target 
experiments and meson factories. Section~\ref{sec:concl} is 
reserved for our summary and conclusions, and some technical 
details are collected in a pair of Appendices.

%%%%%%%%%%%%%%%%%%%%%%%%%%%%%%%%%%%%%%%%%%%%%%%%%%%%%%%%%%%%%%%%%%%%%%

\section{A Warped Hidden Sector\label{sec:rstheory}}

\begin{figure}[ttt]
\begin{center}
%\vspace{1cm}
        \includegraphics[width = 1.0\textwidth]{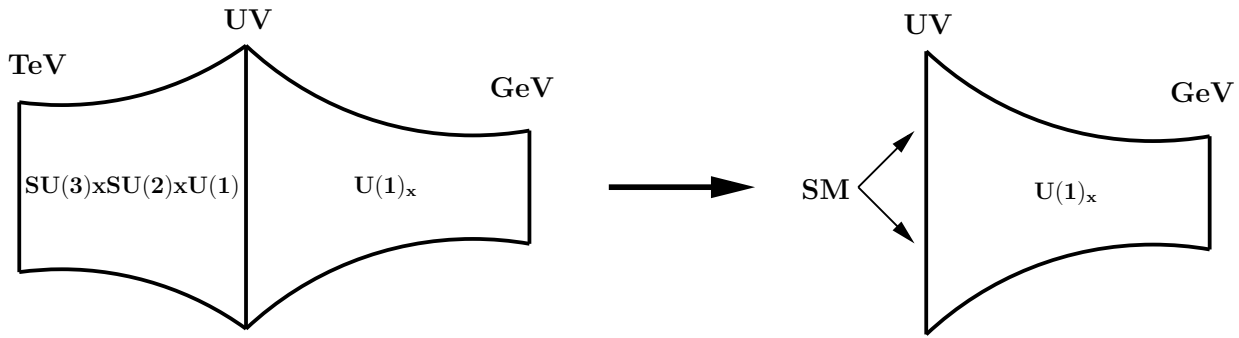}
%\hspace{1cm}
%  \includegraphics[width=0.4\textwidth]{valleyc.eps}
%{./figs_signatures/bgsub.eps}
%\includegraphics[width = 0.35\textwidth]{two_throat_talk.eps}
\end{center}
\caption{Geometry of a warped hidden sector.  The figure on the left
illustrates the two-throat model studied in
Ref.~\cite{McDonald:2010iq}
while the figure on the right shows the simplified construct we investigate
in this work. At energies below the TeV scale the simplified
setup with UV localized Standard Model fields provides a good approximation to the full two-throat model.}%km
\label{throats}
\end{figure}

  In this section we remind the reader of some pertinent features
of the warped hidden sector model investigated in  Ref.~\cite{McDonald:2010iq}
and detail the simplified warped model considered in this work.
The basic setup in Ref.~\cite{McDonald:2010iq} consists of the usual 
RS warped bulk space glued to a second 
``hidden''  warped space at a common UV brane. The IR scale in the visible
``throat'' is near a TeV in order to solve the electroweak hierarchy problem, 
and the usual RS picture is assumed, with an IR localized SM Higgs, bulk SM
fermions and bulk $SU(3)_c\times SU(2)_L\times U(1)_Y$ gauge bosons.\footnote{
For a TeV IR scale in the visible throat to be consistent with precision
electroweak bounds, the gauge symmetry in this throat will likely need to 
be enlarged to include a custodial symmetry as described in 
Ref.~\cite{Agashe:2003zs}. 
However, the low energy physics will still match that described 
here~\cite{McDonald:2010iq}.} 
We take the IR scale in the hidden throat to be at or below a GeV.
A sketch of the model is given in Fig.~\ref{throats}.

%---------------------------------------------------------
%---------------------------------------------------------  

  The low-energy phenomenology of such a multithroat construction is 
quite rich. At energies near the GeV scale the minimal spectrum consists 
of the SM along with a tower of hidden gauge and gravity 
KK modes, whose spacing is on the order of a GeV or less.
The SM couples to the hidden KK vectors primarily via a localized 
kinetic mixing operator connecting $U(1)_x$ and $U(1)_Y$ on the shared UV
brane. The coupling of hidden KK gravitons to the SM is highly
suppressed as the former are strongly localized towards the IR 
of the hidden throat~\cite{McDonald:2010iq}.  Even so, the hidden gravitons 
still play a phenomenologically important role since
their couplings to hidden-sector KK vectors are not suppressed.

  As we are primarily interested in the low-energy phenomenology of 
this setup, it suffices to consider a simplified
picture in which the SM is localized on the UV brane of a single
hidden warped throat which contains a bulk $U(1)_x$ gauge symmetry 
and an IR scale near a GeV (see Fig.~\ref{throats}). 
This picture captures all the important low-energy
physics and we shall employ it in the present work.
At energies above
the TeV scale, the phenomenology of this setup differs from that of
Ref.~\cite{McDonald:2010iq} and one should include the
full two throat structure to accurately determine the high-energy
phenomenology.  
This difference will not significantly modify the
low-energy effects in which we are interested. 
 
  The metric in the 5D bulk space is given by~\cite{Randall:1999ee}
\beq
ds^2= \frac{1}{(kz)^2}(\eta_{\mu\nu}dx^{\mu}dx^{\nu} -
dz^2)= G_{MN} dx^{M}dx^{N},
\label{bulkmetric}
\eeq
where $z \in [k^{-1},\,R]$ labels the extra dimension and $\mu,\nu $
($M,N$) are the 4D (5D) Lorentz indices. As per usual, the Planck mass
is $M_{Pl}^2\simeq M_*^3/k$, where $M_*$ is the 5D Planck scale.
The spectrum of KK gravitons, which we denote as $h_a$, can be found 
by perturbing around the background metric of Eq.~\eqref{bulkmetric}. 
Their masses are~\cite{Davoudiasl:1999jd}
\bea
m_a \simeq \frac{\pi}{R}(a+1/4),\quad a\ge1
\eea
while the massless zero mode $h_0$ is the usual 4D graviton. 

  The UV-localized SM resides at $z=k^{-1}$ along with the following 
gauge kinetic mixing operator~\cite{Holdom:1985ag}
\bea
S\supset -\frac{\epsilon_*}{2\sqrt{M_*}}\,
\int_{UV} d^4x\sqrt{-g }\;g^{\mu\alpha}g^{\nu\beta}B_{\mu\nu}X_{\alpha\beta},
\label{5dcoupling}
\eea
where $g_{\mu\nu}$ is the induced metric, $B_{\mu\nu}$ is the SM hypercharge 
field strength and $X_{\alpha\beta}$ is the $U(1)_x$ field strength. 
Symmetry breaking in the hidden sector can be induced by an explicit
IR-localized Higgs (the ``Higgsed'' case) or by imposing a Dirichlet boundary 
condition on the IR brane 
(the ``Higgsless'' case)~\cite{Csaki:2003dt}.\footnote{For both
sets of boundary conditions we can go to a unitary gauge with $X_5=0$.}  
We shall focus primarily on the Higgsless case in this work, 
but will mention the Higgsed case when there are important differences. 

  For convenience we remind the reader of some essential features of 
the KK spectrum in the Higgsless case. Details of the spectrum and kinetic 
mixing in the Higgsed case can be found in Ref.~\cite{McDonald:2010iq}. 
The KK masses $m_n$ for  the hidden vectors can be approximated as
\beq
m_n \simeq \frac{\pi}{R}(n+1/4)\quad,\quad n>0\ ,
\eeq
and the mass of the lowest mode, which we label as ``0''\!, 
is mildly suppressed relative to the hidden IR scale,
\beq
m_0 \simeq \frac{1}{R}\sqrt{\frac{2}{\log(2kR)-\gamma}},
\eeq
where $\gamma \simeq 0.5772$ is the Euler-Mascheroni constant.
In the effective 4D theory, the UV-localized kinetic mixing operator
induces mixing between SM hypercharge and the tower of KK vectors:
\beq
\mathscr{L}_{eff} \supset -\frac{1}{2}\sum_{n}\epsilon_nX^{\mu\nu}_n
(c_WF_{\mu\nu}-s_WZ_{\mu\nu}),
\eeq
where $s_W$ and $c_W$ refer to the weak mixing angle.
The zero--mode kinetic mixing is given by\footnote{These
  expressions also apply when the SM propagates 
in its own warped bulk after making the replacement
$\epsilon_* \to \epsilon_*\sqrt{k/M_*\log(kR_1)}$.
}
\beq
\epsilon_0 = \frac{\epsilon_*\,f_0(k^{-1})}{M_*^{1/2}} \simeq 
-\epsilon_*\left(\frac{k}{M_*}\right)^{1/2}\frac{1}{\sqrt{\log(2k/m_0)-\gamma}},
\label{epsilonz}
\eeq
where $f_n(k^{-1})$ is the KK mode wavefunction for the $n$-th mode 
on the UV brane.
For the higher modes, one has
\beq
\epsilon_n = \frac{\epsilon_*\,f_n(k^{-1})}{M_*^{1/2}}
\simeq
\,
-\epsilon_*\left(\frac{k}{M_*}\right)^{1/2}\frac{1}{[\log(2k/m_n)-\gamma]}\,(n+1/4)^{-1/2}\ ,
~~~{n \geq 1}.
\label{epsilonn}
\eeq
These expressions show that the kinetic mixing parameter $\epsilon_n$
is mildly suppressed for the higher KK modes relative to the lowest
mode, with $\epsilon_n\simeq \epsilon_0/6\sqrt{n}$ for $n>0$.

Of the five parameters we have introduced to describe the hidden
sector, the gauge coupling will play no role in this work (in the
absence of an IR Higgs). The 5D gravity scale can be fixed by the 4D
Planck mass, leaving three free parameters, which we take as
$\epsilon_0$, $R$ and the
ratio $k/M_*$. As we will detail  below, for energies $E\gg M_*/kR$,
which can be relevant for hidden $Z$ decays and colliders,
the dependence on $R$ drops out of inclusive processes and the
dependence on $k$ is very mild (logarithmic). Absent hierarchically
small values of $k/M_*$ the phenomenology at such energies is
therefore controlled by the single parameter $\epsilon_0$. At lower
energies on the order
of $R^{-1}$ the theory is sensitive to all three free parameters, but
provided $k/M_*$ is non-hierarchical the dependence is primarily on
$\epsilon_0$ and $R$. 

  The application of the AdS/CFT correspondence to RS
models~\cite{ArkaniHamed:2000ds} permits a dual 4D description of
the model we have outlined. In the dual picture there is a
 hidden CFT possessing certain global symmetries, a $U(1)$ subgroup of which
 is weakly gauged. The corresponding gauge boson ($\gamma'$) is a fundamental
 field, external to the CFT, as are the UV localized SM fields
 (see Ref.~\cite{Batell:2007jv} for a discussion of the correspondence
 between KK modes and CFT modes). The
 conformality of the hidden sector is broken explicitly in the UV
 and spontaneously in the IR
 (corresponding respectively to the UV and IR branes in the 5D picture). The 4D
 model contains a gauge kinetic mixing term between $\gamma'$ and SM
 hypercharge ($\sim F_{\mu\nu}'\  B^{\mu\nu}$) and there is further
 kinetic mixing between $\gamma'$ and the spin-one modes of the CFT
 ($\sim F_{\mu\nu}'\ \rho_n^{\mu\nu}$)~\cite{Agashe:2003zs}. The
 latter mixing is akin to the
 kinetic mixing of $\rho$ with the photon in the SM, as electromagnetism
 is a weakly gauged
global symmetry of the QCD sector.
Note that there is no
 direct kinetic mixing term between hypercharge and the CFT modes
 $\rho_n$, but one is induced by their common mixing with
 $\gamma'$. The SM therefore couples to the CFT modes by its weak
 coupling (for $E\ll k$) to the fundamental field $\gamma'$.

  Before proceeding we note that the hidden sector could also contain
additional states with non-zero $U(1)_x$ charge, possibly including 
a UV-localized dark matter candidate as mentioned in~\cite{McDonald:2010iq},
or new matter fields in the bulk. 
New states on the UV brane will not significantly modify the
low-energy phenomenology we discuss provided they are at or above
the weak scale.  Indeed, 
the (unspecified) dynamics on the UV brane
that ensures the stability of the electroweak sector can also lead
to new states in the electroweak mass range.  Exotic hidden bulk matter will 
naturally be much lighter, with zero-mode and low-lying KK mode masses near 
the characteristic mass scale of the hidden IR brane.  This could, for example,
lead to a relatively light dark matter candidate with a mass of order
MeV--GeV.  Such new light states could significantly
modify the low-energy phenomenology relative to the minimal model
we consider here; if they are lighter (heavier) than the vector zero mode 
predominantly (partially) invisible final states can arise.
We defer the study of this possibility to 
a future work.

  We also note that previous works have considered warped models
with sub-weak scales~\cite{Gripaios:2006dc,McDonald:2010jm}, and that
much of the phenomenology we consider here would be similar if
implemented on the truncated space of Ref.~\cite{McDonald:2010jm}. 
The gauge kinetic mixing we consider
can also find its origin in string models~\cite{Abel:2008ai} and
our phenomenological analysis may be of interest in this regard.

%%%%%%%%%%%%%%%%%%%%%%%%%%%%%%%%%%%%%%%%%%%%%%%%%%%%%%%%%%%%%%%%%%%%%%

\section{Hidden Sector Decays\label{sec:decay}}

Kinetic mixing between $U(1)_x$ and hypercharge permits the creation
of hidden KK vectors in 
experiments colliding SM fields together. 
The expected experimental signal once a hidden vector is 
produced depends
on the decay properties of the vector.  It is therefore important to
understand these properties to determine whether the vectors 
decay predominantly into the hidden sector or to the SM, and to determine 
likely signals. In this section we consider the
decay properties of the KK vectors in some detail. 
Related discussions in a different context can be found
in Refs.~\cite{Csaki:2008dt,Stephanov:2007ry,Strassler:2008bv}.

  The decay of a hidden KK vector $X_n$ to SM fields requires a kinetic
mixing insertion and the corresponding widths are suppressed by a
factor of $\epsilon_n^2\ll1$. For example, the decay width of the
$n$-th KK mode to a pair of SM leptons $\ell\bar{\ell}$ 
is~\cite{Pospelov:2008zw}
\bea
\Gamma(X_n\rightarrow \ell\bar{\ell})=
\frac{1}{12\pi}e^2 c_W^2\epsilon_n^2m_n\left(1+\frac{2m_\ell^2}{m_n^2}\right)
\left(1-\frac{m_\ell^2}{m_n^2}\right)^{1/2},
\eea
for $m_n\ll m_Z$, and similarly the width to hadrons is
\bea
\Gamma(X_n\rightarrow \mathrm{hadrons})
~=~\Gamma(X_n\rightarrow \mu^+\mu^-)\,R(s=m_n^2),
\eea
where $R(s)$ is the usual hadronic $R$ parameter, 
\mbox{$R=\sigma(e^+e^-\rightarrow
  \mathrm{hadrons})/\sigma(e^+e^-\rightarrow \mu^+\mu^-)$}~\cite{Amsler:2008zzb,Ezhela:2003pp}. 
As the lowest data point in the hadronic cross section data 
set is at $\sqrt{s}=0.36$~GeV, well above the pion threshold, we
follow~\cite{Batell:2009yf} and use the cross section for
$e^+e^-\rightarrow\pi^+\pi^-$ in the uncharted region above 
threshold~\cite{Ezhela:2003pp,Davier:2002dy}. Observe that the total
decay width to the SM is roughly independent of mode number, up to a
growth in the number of kinematically accessible states, as
the KK mass grows like $m_n\sim n$ while $\epsilon_n^2$ goes 
like $\sim 1/n$.  These decays are relatively prompt on collider 
timescales ($c\tau < 1\,mm$) for $R^{-1} \sim \gev$ provided 
$\epsilon_0 \gtrsim 10^{-4}$~\cite{Reece:2009un}.

  Even in the absence of hidden sector matter, 
heavier KK vectors can 
still decay within the hidden sector via the creation of a KK graviton 
and a lighter vector mode, 
$X_n\rightarrow h_aX_m$. These decays are kinematically allowed for 
sets of KK numbers satisfying $n>m+a$ in both the Higgsed and Higgsless 
cases, but decays with $a=0$ vanish due to wavefunction orthogonality. 
Therefore all vector modes with $n\ge 2$ can decay via graviton production. 
We present a computation of the decay width for $X_n\rightarrow h_aX_m$ 
in Appendix~\ref{app:vector_decay}.

\begin{figure}[ttt] 
\centering
\includegraphics[width=0.55\textwidth]{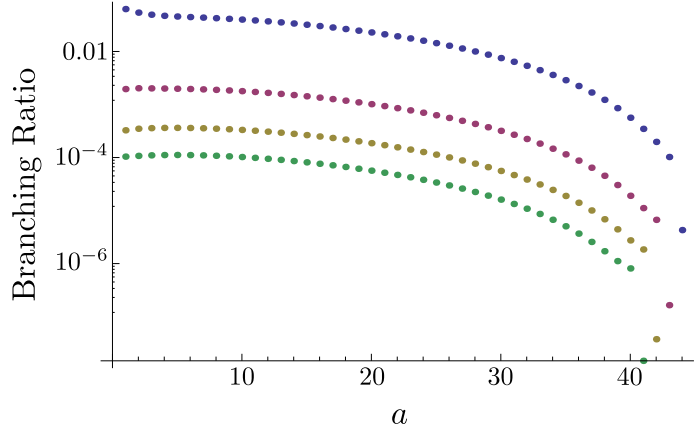} 
  \caption{Plot of the Branching Ratio Vs. Daughter KK
    number $a$ for the
    hidden sector decay of the $n=45$-$th$ KK vector. We plot 
    fixed values of $(a+m)$ and from top
    to bottom the curves satisfy $(a+m)=(44,43,42,41)$. The branching
    ratio is taken as $BR\simeq\Gamma(X_{n=45}\rightarrow h_a
    X_m)/\sum_{m,a}\Gamma(X_{n=45}\rightarrow h_a X_m)$ and we use
    $R^{-1}=1$~GeV. The plot is for the Higgsed Case but similar
    behavior is found in the Higgsless case.} 
  \label{fig:vector_n_45_width_v_daughterkk_number_higgs}
\end{figure}

  For a given KK level $n$, the partial width $\Gamma(X_n\rightarrow h_a
X_m)$ can vary substantially as one varies the daughter KK numbers $a$
and $m$. To demonstrate this we plot the branching ratio for the decay
$X_n\rightarrow h_a X_m$ against daughter KK number $a$ in
Fig.~\ref{fig:vector_n_45_width_v_daughterkk_number_higgs} for the
mode $n=45$. The Figure shows points with fixed values of
$m+a= (44,43,42,41)$. Decays with values of $a+m<41$ are subdominant
to those plotted. The plot reveals some important features. Firstly,
one observes that decays with daughters $h_a$ and $X_m$ satisfying
$a+m\sim n$ are dominant.  For a fixed value of the graviton mode number
$a$ one can see that the branching ratio for decays with $a+m=44$ is 
more than a hundred times larger than that with $a+m=42$. This disparity 
increases as one decreases the sum $a+m$. Also note that for 
fixed values of $a+m$ the
decays into lighter KK gravitons dominate.

%----------------------------------------------------%
%----------------------------------------------------%

  The tendency of KK vectors to decay into daughters satisfying $n\sim
a+m$ demonstrates an approximate conservation of KK number present in
RS models~\cite{Csaki:2008dt}. 
In the absence of warping, 
the momentum along the extra
dimension would be conserved and KK decays would necessarily conserve
KK number with $n=a+m$. Turning on the
warping breaks the translational invariance along the extra dimension 
so that KK number is no longer conserved,
however an approximate KK number conservation
persists and is encoded in the wavefunction overlap factors
$\zeta_{a,mn}$ and $\xi_{a,mn}$ presented in
Appendix~\ref{app:vector_decay}. As
different KK modes are localized
at different points along the extra dimension the overlap factors 
contain oscillatory integrands unless $n\sim a+m$. This results in
a suppression of $\zeta_{a,mn}$ and $\xi_{a,mn}$ as one increases the
separation between $n$ and $a+m$, explaining the dominance of decays
with $n\sim a+m$. 

%----------------------------------------------------------------------%
\begin{figure}[t] 
\centering
\includegraphics[width=0.55\textwidth]{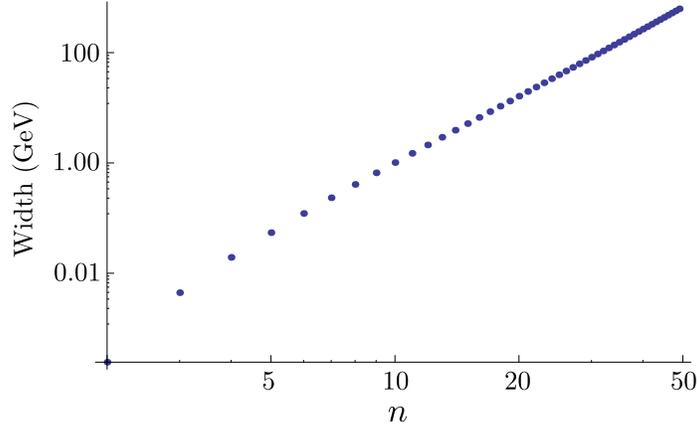} 
  \caption{Plot of $\sum_{a,m}\Gamma(X_n\rightarrow h_a X_m)$ Vs. KK number $n$, where the sum is over values of $a,m$
   satisfying $n>a+m$. The plot is for the Higgsless Case but similar
   behavior is observed in the weakly
   Higgsed case.}
  \label{fig:total_width_v_kk_number_50_nohiggs}
\end{figure}
%---------------------------------------------------%
  
  In Fig.~\ref{fig:total_width_v_kk_number_50_nohiggs}
we show the total decay width of the $n$-th mode
due to two-body decays to graviton and gauge modes for $n \le 50$.  
The plot is for the Higgsless case with  $k/M_*=0.1$ and $R^{-1}=1$~GeV.
One observes that a simple parameterization of the total hidden two-body decay 
width is possible for $n \gg 1$:
\bea
%\Gamma(X_{n}\rightarrow Hidden)=
\Gamma_n \equiv
\sum_{m,a}\Gamma(X_{n}\rightarrow h_a X_m)
=\frac{k^2}{M_{Pl}^2}\frac{g_*(n)}{8\pi}m_n\ ,\label{width_param}
\eea 
where the effective coupling constant $g_*(n)$ can be written as
\bea
g_*(n)= \mathcal{C} n^p,
\eea
with $\mathcal{C}\simeq0.80$ and $p\simeq3$ for $R^{-1}=1$~GeV.
%\bea
%\mathcal{C} &\simeq& \left\{\begin{array}{cl}0.80& \mathrm{Higgsless}\\ 0.64& \mathrm{Higgsed}\end{array}\right.,\\
% p&\simeq& \left\{\begin{array}{cl}3.00& \mathrm{Higgsless}\\ 2.97& \mathrm{Higgsed}\end{array}\right.,
%\eea
Numerically we find that
varying $R$ does not 
appear to
change the power $p$ but does induce a small change in the constant 
$\mathcal{C}$. For example, varying $R$ over an order of magnitude 
produces a shift in $\mathcal{C}$ of order a few to ten percent.
It will also be helpful to have a parametrization for the hidden width 
of the lighter modes, $n\sim 1$. Writing these as in 
Eq.~(\ref{width_param}) with
 $\mathcal{C}\rightarrow\mathcal{C}_n$ we find
\bea
\mathcal{C}_2\simeq 0.10\quad,\quad \mathcal{C}_3\simeq
0.40\quad,\quad \mathcal{C}_4\simeq 0.58\quad,\quad
\mathcal{C}_5\simeq 0.66\ ,
\eea
for the Higgsless case with $R^{-1}\lesssim 1$~GeV. 
The parameters $\mathcal{C}_n$ vary slowly as
one varies the length of the extra dimension $R$,
but the
above values 
provide a reasonable approximation for the energies of interest to us.

   The parametrization for the hidden two-body decay width $\Gamma_n$ in
 Eq.~\eqref{width_param} 
gives a good approximation to the total KK mode width
when the KK description is perturbatively under control. 
However, the KK description breaks down at energies of order $(M_*/k)R^{-1}$ 
as the coupling between KK modes becomes strong~\cite{Goldberger:2002cz}. 
This is borne out in Eq.~\eqref{width_param}, which shows that $\Gamma_n$ 
is parametrically of the same order as $m_n$ for $n \sim (M_*/k)$,
indicating that the modes are bleeding into each other and the KK
description is breaking down.  
At strong coupling higher-order corrections and multi-body decays are 
expected to become important.  Decays to stringy modes, 
either from a string theoretic UV completion or from strong coupling, 
may also become 
significant~\cite{Strassler:2008bv,Reece:2010xj}. 
We will discuss a method for calculating 
inclusive decay widths
at higher energies in Sec.~\ref{sec:match}.

  To illustrate the decay channels of the calculable lower KK modes
we consider the Higgsless case for
$k/M_*=0.1$ and $R=1\,\gev^{-1}$ with 
the strength of the zero-mode kinetic mixing fixed at 
$\epsilon_0=3\times 10^{-3}$.  Once these parameters are set 
the kinetic mixing for
the higher modes is determined by Eq.~(\ref{epsilonn}) as 
$\epsilon_n\simeq \epsilon_0/6\sqrt{n}$. In this case the zero mode 
has mass $m_0\simeq 230$~MeV and can only decay leptonically,
$X_0\rightarrow 2e,2\mu$, with a width of about a keV. The first KK mode 
($n=1$) also decays to light SM leptons and hadrons with a slight 
preference for leptons, $BR^{(1)}_{\mathrm{lep}}=0.56$,
and a total width of about a keV. The second KK mode ($n=2$) can
decay within the hidden sector in addition to the SM 
and has a much larger width on the order of $0.2\,\mev$.  
Decays to the hidden sector almost completely dominate with a branching 
ratio $BR(X_2\rightarrow X_0h_1)=0.9998$.  Of the decays to the SM, 
hadronic 
modes
are slightly preferred with 
$BR^{(2)}_{\mathrm{had}}=2.9\times 10^{-4}$ and
$BR^{(2)}_{\mathrm{lep}}=2.3\times 10^{-4}$. As the width to the SM
does not significantly increase for heavier modes, the $n\geq 2$ modes
all decay predominantly within the hidden sector. 

  Once created, a KK graviton can decay to lighter KK vectors. 
We present the two-body width for graviton decay $h_a\rightarrow
X_mX_n$ in Appendix~\ref{app:decay_kk_grav},
where we show that graviton decays further demonstrate the
approximate conservation of KK number and prefer decays with $a\sim
m+n$. These decays are prompt, as the coupling between KK vectors and
gravitons is set by $R^{-1}$.  For example, in the Higgsless case with 
$k/M_*=0.1$ and $R^{-1}=\gev$ we find 
$\Gamma(h_1\rightarrow 2X_0)\simeq 3\times 10^{-2}$~MeV for the
first KK mode. The total width for the heavier vectors increases with
KK number $a$ due to the increase in  kinematically available final states 
(see Appendix~\ref{app:decay_kk_grav}). 
Heavier KK gravitons can also decay to lighter KK gravitons.
Note, however, that the lightest KK graviton  ($a=1$)  will decay 
entirely to pairs of $X_0$.

  Putting these pieces together, there emerges a simple picture 
of decays within a Higgsless hidden sector.  The $n=0,\,1$ vector KK modes 
decay directly back to the SM once they are produced.  
When a higher vector KK mode (or a superposition of them) 
is created, such as in an energetic collision of SM fields,
it decays promptly to lighter KK vector and graviton modes.
In turn these decay to even lighter hidden states, 
producing a cascade down the hidden KK tower. This showering 
terminates at the $n=0,\,1$ vectors which decay back to the SM. 
Therefore the generic final state resulting from the production of 
heavier hidden sector modes is a high multiplicity of relatively soft 
and light SM particles, similar to the scenario discussed
in Ref.~\cite{Strassler:2008bv}.

  The presence of an IR-brane hidden Higgs field $h_x$ can 
significantly modify the experimental signals of the hidden cascade.
With such a Higgs, KK vectors can decay through the process 
$X_n\rightarrow X_m +h_x$ when  kinematically available.  
The precise way in which the decay spectrum is altered depends on the
relationship between the Higgs mass 
and the vector KK masses (similar to the 4D
case~\cite{Batell:2009yf,Gopalakrishna:2008dv,Schuster:2009au}). 
If $m_h$ lies below twice the lightest KK vector mass $m_0$, 
the hidden Higgs will decay only slowly back to the SM through 
multi-body or loop-induced channels involving $X_0$.  
These decays are typically slow relative to experimental timescales 
for light hidden Higgs masses 
leading to missing energy in the final state for $m_h< m_0$,
and possibly displaced vertices for $m_0 < m_h < 2m_0$~\cite{Batell:2009yf}.  
A light Higgs will also allow $X_1\to X_0+h_x$, which almost always
dominates over direct decays to the SM and can produce 
four- and six-lepton final states.  
As $m_h$ grows larger than $(m_1-m_0)$ the $n=0,\,1$ vector modes
again decay exclusively to the SM.  The hidden Higgs can still be produced 
in the decays of heavier KK modes and will decay promptly to pairs of 
lighter vectors.  Hidden decay cascades in this case will be qualitatively
very similar to the Higgsless case described above.  

   Vector and graviton KK modes may also have decays involving a bulk
radion $r_x$, such as $X_n\rightarrow X_m+r_x$.  
These will be very similar to decays involving an IR-brane 
hidden Higgs with which it could mix~\cite{Csaki:2000zn}.  
As for a hidden Higgs, if the radion is very light it can 
radically alter the decays of the lightest modes, but as its mass increases 
it mainly affects the heavier KK modes.  The precise determination of the 
radion mass is a 
somewhat 
complicated subject that requires a specification of the
stabilization dynamics and a determination of their backreaction on
the metric~\cite{DeWolfe:1999cp}. While there is no difficulty of principle 
in stabilizing the hidden warped extra
dimension~\cite{Goldberger:1999uk} (even in the full two-throat
setup~\cite{Law:2010pv}), 
in practise the precise value of the radion mass will depend on the details
of the approach adopted. Provided the radion is heavier than the $n=0$
and $n=1$ 
vector KK modes it will not significantly modify the cascade picture
presented above. We assume this to be the case in the present work 
and neglect radion decay channels.

  In summary, we find that the dominant two-body decay of the $n$-th
KK vector is into light KK gravitons $h_a$ and KK vectors $X_m$ with
$m\sim n$ and $n\sim a+m$ for $n\ge 2$. If kinematically
permitted the daughter vector will further decay into lighter
gravitons and vectors and the daughter gravitons will themselves 
decay back into two lighter vectors.  The creation of heavier modes 
therefore results in a cascade decay down the tower in the hidden sector 
until one ends up with a collection of light KK vectors. 
The light vectors then decay to light SM fields.

%%%%%%%%%%%%%%%%%%%%%%%%%%%%%%%%%%%%%%%%%%%%%%%%%%%%%%%%%%%%%%%%%%%%%%%%%%%%%%

\section{Matching to Five Dimensions\label{sec:match}}

  In the phenomenological analyses to follow, we will be interested
in processes initiated by UV-localized SM states at energies both 
well above and well below the hidden IR scale $R^{-1} \sim \gev$,
but always much less than the UV scale $k$. Since the local cutoff at a
given point $z$ along the extra dimension is $M_*/(kz)$, 
the effective theory description breaks down in the
IR for energies approaching 
$(M_*/k)R^{-1}$~\cite{ArkaniHamed:2000ds,Goldberger:2002cz}. 
For processes initiated by SM fields on the UV brane, however,
where the local cutoff is near the Planck scale, one expects a 
reliable effective theory description 
to exist~\cite{ArkaniHamed:2000ds,Goldberger:2002cz,Pomarol:2000hp}.
At energies below $(M_*/k)R^{-1}$ only the lightest KK
modes are important and the KK effective theory developed above 
provides a reliable description of the full dynamics.  
For energies greater than this, the KK modes become increasingly broad and
strongly-coupled, and the KK description breaks down. 
However, 
the full 5D theory remains weakly-coupled up to the UV scale 
for processes initiated on the UV brane~\cite{Goldberger:2002cz}.
We make use of this fact by matching the 4D KK theory 
onto the 5D theory at energies near $R^{-1}$, where both descriptions 
are sensible.  Our matched 5D theory then allows us to compute reliably 
in the intermediate regime $R^{-1} \lesssim E \ll k$ for processes initiated
on the UV brane. In this section we detail this matching. We note that the
technical details of this section, though important for our analysis,
lie outside of our main
phenomenological focus. Readers not concerned with these
computational 
details
can proceed directly to Sec.~\ref{sec:pew}.

  Consider first the 5D bulk vector propagator at tree-level.  
We will always contract this propagator with a transverse projector 
$(-p^2\eta_{\mu\nu}+p_{\mu}p_{\nu})$ arising from the kinetic mixing 
interaction of Eq.~\eqref{5dcoupling},
so we only need the gauge-invariant coefficient of
the transverse portion of the propagator.  With Higgsless 
(Neumann-Dirichlet) or Higgsed (Neumann-Neumann) boundary conditions 
it is given by
\bea
\Delta_p(z,z') &=& 
\left\{
\begin{array}{cc}
{
\frac{\pi}{2}\frac{kzz'}{(Y_{1,R}J_{0,k}-J_{1,R}Y_{0,k})}
}
(J_{1,R}Y_{1,>}-Y_{1,R}J_{1,>})(J_{0,k}Y_{1,<}-Y_{0,k}J_{1,<})
&
\text{(Higgsless)}
%\label{higgsless_prop}
\\
\phantom{\frac{I}{I}}&
\\
\frac{\pi}{2}\frac{kzz'}{(Y_{0,R}J_{0,k}-J_{0,R}Y_{0,k})}
(J_{0,R}Y_{1,>}-Y_{1,R}J_{1,>})(J_{0,k}Y_{1,<}-Y_{0,k}J_{1,<})
&
\text{(Higgsed)}
%\label{higgsed_prop}
\end{array}\right.
\label{vec_prop}
%&&
% \frac{k}{p^2[\log(2k/p)-\gamma]}\left(1
%+\frac{\pi}{2}\tan(p/T-{3\pi}/{4})/
%[\log(2k/p)-\gamma]\right)^{-1},
%~~~~~z=z'=k^{-1},~T \ll p \ll k,
%\nnmb
\eea
where $p = \sqrt{p^2}$, $z_> = max\{z,z'\}$, 
$z_< = min\{z,z'\}$ and $J_{n,z}$ is shorthand for $J_n(pz)$. The
Neumann-Neumann propagator was derived in~\cite{Randall:2001gb}.
Taking $z = z' = k^{-1}$ and $R^{-1} \ll |p| \ll k$, 
both 5D tree-level propagators reduce to
\beq
\Delta_p(k^{-1},k^{-1}) \simeq \frac{k}{p^2[\log(2k/p)-\gamma]}\times \left\{
\begin{array}{ccc}
 \left(1+\frac{\pi}{2}\tan(pR-{\mathcal{N}\pi}/{4})/
[\log(2k/p)-\gamma]\right)^{-1}
%&;&  p^2>0
\\
\phantom{\frac{I}{I}}\label{uv2uv}
\\
-1 
\end{array}\right.
\eeq
where the upper (lower) term is for $ p^2>0$ ($p^2<0$) and
$\mathcal{N} = 3~(1)$ for the Higgsless (Higgsed) case.
Both expressions have poles corresponding to the KK masses.
In fact it can be shown that the tree-level 5D
propagators are equal to a sum over KK modes weighted by
KK bulk wavefunction factors $f_n(z)$,
\beq
\Delta_p(z,z') = \sum_n\frac{f_n(z)f_n(z')}{p^2-m_n^2},
\eeq
where the masses $m_n$ are given in Section~\ref{sec:rstheory}.\footnote{
Explicit expressions for the KK wavefunction factors $f_n(z)$ can
be found in Ref.~\cite{McDonald:2010iq}.} 

  The tree-level propagators in Eq.~\eqref{uv2uv} will be modified 
in an important way by quantum effects~\cite{Strassler:2008bv,Goldberger:2002cz,Pomarol:2000hp,Randall:2001gb,Choi:2002wx,Agashe:2002bx}.
In addition to mass and wavefunction corrections, 
the RS1 propagators will acquire complex
self-energies for $p \gtrsim R^{-1}$.  These arise from bulk gauge-graviton 
(or gauge-Higgs) loops, and come with factors of $k/M_*$. 
Rather than carrying out an explicit 5D calculation of the 
loop corrections to the gauge boson self-energy, we take a different
approach based on matching to RS2~\cite{Randall:1999vf},
where the IR brane is taken to $z\to \infty$, 
which we argue is a universal IR-independent limit of the 
RS1 propagators at high momentum~\cite{ArkaniHamed:2000ds}. 

  For reference, the coefficient of the gauge-invariant transverse 
part of the RS2 UV-to-UV propagator is given 
by~\cite{Dubovsky:2000am,Friedland:2009iy}
\bea
\Delta_p^{RS2}(k^{-1},k^{-1}) &=& \frac{H_1^{(1)}(p/k)}{pH_0^{(1)}(p/k)}
\label{rs2prop}
\\
&\simeq& 
\frac{k}{p^2[\log(2k/p)-\gamma]}\times
\left\{
\begin{array}{ccc}
\left(1
-i\frac{\pi}{2}/[\log(2k/p)-\gamma]\right)&;&p^2>0\\
\phantom{i}&\\
- 1
&;&p^2<0
\end{array}
\right.
\nnmb
\eea
where $H_n^{(1)} = (J_n+iY_n)$.
Note the appearance of an imaginary part for $p^2>0$, which
follows from the imposition of
outgoing-wave boundary conditions as 
$z\to \infty$~\cite{Dubovsky:2000am,Giddings:2000mu},
and represents the escape of vectors into the bulk. 

  Comparing the \emph{tree-level} RS1 UV-to-UV propagators 
in Eq.~\eqref{uv2uv} to the RS2 expression of Eq.~\eqref{rs2prop},
we see that they agree at spacelike momentum with
$|p|>R^{-1}$.
At large timelike momentum, the real parts of the RS1 propagators also
match closely with RS2, except near the poles of the log-suppressed 
tangent term.  This agreement can be understood in terms of the
UV-to-bulk RS propagators, which become highly oscillatory or 
exponentially damped for $z > |p|^{-1}$.  Thus, the full theory 
appears to be insensitive to the detailed geometry at $z \gg |p|^{-1}$.  
In particular, modifying the geometry at large $z$ by adding an IR brane 
or changing its location should have virtually no effect on the 
quantum-corrected UV-to-UV propagator at large 
momentum~\cite{ArkaniHamed:2000ds}.
 
  There is a potential loophole in this argument related
to the poles of the tree-level RS1 propagators.  These poles correspond 
to KK masses, and reflect 
an apparent
sensitivity to the IR that persists at 
large momentum.  The tree-level RS1 propagators also lack the imaginary 
term present in the RS2 propagator for timelike momentum.
Both of these features are artifacts of the tree-level 
approximation wherein the KK modes are treated as being absolutely stable. 
Consequently, the KK poles at this level may be thought of as standing waves 
that reflect off the IR boundary, thereby probing the deep IR and ensuring 
there is no net dissipation from the UV brane.  

  Quantum corrections to the propagator will generate an 
imaginary component in the self-energy.  Near the pole of a narrow
low-lying KK mode we can identify this imaginary piece with the finite 
decay width computed in Sec.~\ref{sec:decay}. Its effect is to regulate the KK
pole divergences in the tree-level RS1 propagator, and it represents a net 
dissipation from the UV brane.  For lighter modes that are very narrow,
some probability for reflection remains and the net dissipation
differs to that of RS2.  Going to higher timelike momentum will probe 
heavier KK modes that are increasingly broad.
When the widths become as large as the mode spacing $\pi/R$,
the KK resonances begin to overlap significantly and 
approach a continuum~\cite{Strassler:2008bv}.  
We expect this continuum to be independent of the IR structure of the 
theory, and to persist to even higher momenta where the KK theory becomes 
strongly-coupled.
Intuitively, a broad KK mode produced on the UV brane 
will decay (or shower) well before reaching the IR brane and can
no longer probe the deep IR.  As the probability of this mode reflecting 
off the IR brane falls to zero, the dissipation due to decays should 
agree with RS2 where no reflection is possible.  

  We therefore argue that the quantum-corrected 5D
RS1 UV-to-UV propagator in the momentum range
$(M_*/k)R^{-1} \ll |p| \ll k$ approaches a universal IR-independent 
limit that is approximated well by the \emph{tree-level} 
RS2 UV-to-UV propagator. 
These properties are already true at tree-level for spacelike 
momentum with  $|p| \gg R^{-1}$.  
At large timelike momentum, perturbatively small quantum corrections 
smooth out the poles and generate an imaginary component in the propagator,
producing a continuum that is insensitive to the IR geometry. 
Since the 5D RS theory is perturbatively calculable for processes
initiated on the UV brane with $|p| \ll k$, quantum corrections to the  
the propagator are numerically small.  In the case of RS2,
the tree-level propagator is smooth at timelike momentum, and can not
be modified strongly by perturbative loop effects.  The universal 
high-momentum limit should thus be described well by the tree-level 
RS2 propagator.

  This matching is also consistent with the 
gauge dual picture of a 4D 
approximate CFT.  The presence of an IR brane corresponds to a spontaneous
breaking of conformal invariance at low energies.  Going to energies well above
the breaking scale, CFT-breaking effects will become increasingly irrelevant,
falling off as positive powers of $R^{-1}/E$, and the theory will approach
a pure CFT (with a UV cutoff) 
that coincides with RS2.  
%5
Let us also point out that the 5D quantum corrections we have
considered correspond to subleading corrections in a $1/N$ expansion 
within the CFT dual.

%----------------------------------------------------------------------%
\begin{figure}[ttt] 
\centering
\includegraphics[width=0.5\textwidth]{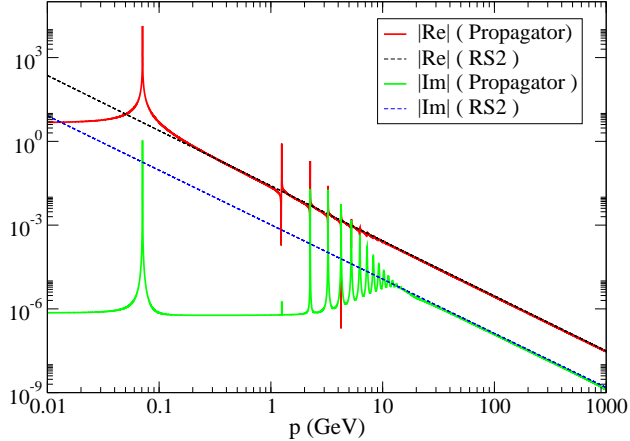} 
  \caption{Real and Imaginary parts the 5D UV-to-UV propagator
of Eq.~\eqref{prop_emp} together with those for RS2.  We use
$k/M_* =0.1$ and $\pi/R = 1\,\gev$. }
  \label{fig:props}
\end{figure}
%---------------------------------------------------%

  Our strategy for the rest of the paper is to match the KK theory valid 
at low energies to the RS2-like 5D limit we argue is valid at higher energies.  
For this, we use the following \emph{empirical} form for the 
quantum-corrected UV-to-UV 5D propagator
to interpolate between the two regimes:
\beq
\Delta_p^{UV} = \sum_n\frac{f_n^2(k^{-1})}{p^2-m_n^2+ip\tilde{\Gamma}_n},
\label{prop_emp}
\eeq
where
\beq
\tilde{\Gamma}_n = \left\{
\begin{array}{cc}
\Gamma_n\,;&~~~~~\Gamma_n \leq m_n/4\\
m_n/4\,;&~~~~~\Gamma_n > m_n/4
\end{array}
\right.,
\eeq
and $f_n(k^{-1}) = \epsilon_n\,(M_*^{1/2}/\epsilon_*)$.
This form reproduces the correct KK pole structure at low-momenta where
the KK theory is calculable and the resonances are narrow, 
and connects smoothly to the RS2 propagator for $|p| > (M_*/k)R^{-1}$
over the range of momenta of interest.\footnote{{This expression
does not fully account for the pole structure in the intermediate
regime when the KK modes have widths on the order of the mass 
spacing $\pi/R$~\cite{Cacciapaglia:2009ic}, 
but we do not expect this to modify our results significantly.}}
In Fig.~\ref{fig:props} we plot both the real and
imaginary parts of Eq.~\eqref{prop_emp}, together with those of
the RS2 propagator given in Eq.~\eqref{rs2prop}.  At large momenta
the expression of Eq.~\eqref{prop_emp} is relatively insensitive to the
precise form of $\tilde{\Gamma}_n$ provided it is larger than the
KK mode spacing.  Fig.~\ref{fig:props} also illustrates how the finite
KK mode widths smooth out the poles in the tree-level propagator.
Although we do not show it here, the form of Eq.~\eqref{prop_emp}
also agrees well with the RS2 propagator for $p^2 <0$ 
when $|p|\gg R^{-1}$.

%%%%%%%%%%%%%%%%%%%%%%%%%%%%%%%%%%%%%%%%%%%%%%%%%%%%%%%%%%%%%%%%%%%%%%

\section{Precision Electroweak Observables\label{sec:pew}}

  Having developed the necessary tools we now proceed to the detailed
phenomenology. We will consider precision
electroweak observables in this section and  turn to low-energy signals
and constraints in Sec.~\ref{sec:lowe}. For simplicity we focus
primarily on the Higgsless case with a Dirichlet IR boundary
condition in both sections.  The Higgsed case will be 
qualitatively similar, with the main difference being the presence of
additional decay channels (like Higgs$'$-strahlung) that can modify 
the signals in low-energy experiments.

\begin{figure}[ttt]
\begin{center}
%\vspace{1cm}
        \includegraphics[width = 0.30\textwidth]{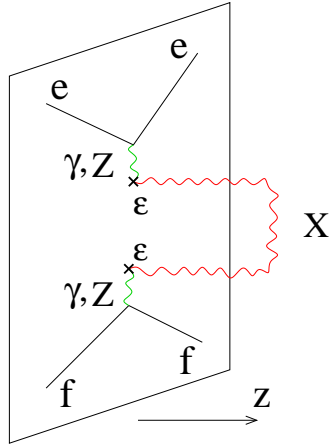}
\end{center}
\caption{Correction to the $Z$ propagator. The hidden vector
  is created on the UV brane (represented by the
  plane) but may propagate into the bulk before producing standard 
model states.}
\label{bulk}
\end{figure}

  Mixing between the hidden bulk gauge vector and the SM photon 
and $Z$ will alter the predictions for electroweak observables.  
The two classes of observables we study in this section are
$e^+e^-\to f\bar{f}$ scattering at $\sqrt{s} \sim m_z$, 
where $f$ is a SM fermion, as well as rare $Z$ decay modes.
To compute these effects we treat the kinetic mixing operator of 
Eq.~\eqref{5dcoupling} as an interaction that couples the SM to the 
bulk vector.  For $e^+e^-\to f\bar{f}$ we only need the UV-to-UV
bulk propagator, for which we use the matched expression
given in Eq.~\eqref{prop_emp}.  This is illustrated schematically 
in Fig.~\ref{bulk}.
By way of a unitarity cut, we can also
use this propagator to compute the inclusive rate of rare $Z$ decays.  
The direct production of hidden sector states at lower energies will
be considered in Sec.~\ref{sec:lowe}.

\subsection{$\mathbf{e^+e^-\to f\bar{f}}$ Processes\label{pew_e-->f}}

%---------------------------------------------------------
%---------------------------------------------------------  

  The interaction vertices for $X$-SM mixing in unitary gauge are
\bea
Z X:&&~~~~~-i\frac{\epsilon*}{M_*^{1/2}}s_W\,p^2
\left(-\eta_{\mu\nu}+p_{\mu}p_{\nu}/p^2\right)
\; \delta(z-k^{-1})\ ,
\\
\gamma X:&&~~~~~\phantom{-i}i\frac{\epsilon*}{M_*^{1/2}}c_W\,p^2
\left(-\eta_{\mu\nu}+p_{\mu}p_{\nu}/p^2\right)
\; \delta(z-k^{-1})\ .
\eea  
Notice that these involve transverse projectors -- contracting $p_{\mu}$
into the vertices gives zero. We can apply these vertices to the
process $e^+e^-\to
f\bar{f}$ with $f\neq e$, working to leading non-trivial
order in $\epsilon_*$ and neglecting 
fermion masses.  With the resulting summed and squared matrix element,
which is given in Appendix~\ref{app:pew},
we have everything
we need to compute 
the full set of precision electroweak observables and compare to experimental
observations.  

  For this, we will take $m_Z$, $G_F$, and $\alpha_{em}(Q^2\to 0)$
as input parameters for computing other electroweak observables.
The effect of a tower of light hidden KK vectors on 
$G_F$ and $\alpha_{em}(Q^2\to 0)$ is dominated by the lightest
mode.
The Fermi constant 
$G_F$ is measured in muon decay, and 
is not changed at all at leading order.\footnote{We work implicitly
to leading order in $\epsilon$ effects, and drop any $\epsilon$
effects appearing at loop order when the main contribution is tree-level.}    
Currently, the best-measured value of $\alpha_{em}(Q^2\to 0)$ comes
from $(g-2)_e$.  This can be shifted by a light hidden vector,
but the size of the shift cannot be so big as to disagree with the
value of $\alpha_{em}(Q^2\to 0)$ measured in atomic systems, which are
not significantly modified by the new states~\cite{Pospelov:2008zw}.
Both determinations have a much higher precision than the value 
of $\alpha$ extrapolated to the weak scale, which has sizeable
hadronic uncertainties.  Thus, once the low-energy bounds are satisfied,
the effective shift in $\alpha(m_Z)$ relevant for precision electroweak 
observables is negligible.

  Determining the value of $m_Z$ is more complicated.  It is obtained from
a parametrized fit (with $m_Z$ as a parameter) of $e^+e^-\to f\bar{f}$
cross-sections measured over a range of energies near the $Z$ pole. 
Hidden vectors will modify this $Z$ lineshape, and can thereby
lead to a poor fit or a shift in the extracted value of $m_Z$.
To investigate these effects we fix $\epsilon_*$ such that 
$|\epsilon_0| = 1\times 10^{-2}$, set $k/M_* = 0.1$
and $R = \pi/\gev$, and compute
the resulting $e^+e^-\to f\bar{f}$ cross-sections.
Note that for masses of the lightest hidden vector below $10\,\gev$,
$|\epsilon_0| = 10^{-2}$ is slightly larger than what is consistent 
with low-energy probes~\cite{Bjorken:2009mm,Pospelov:2008zw}.

\begin{figure}[ttt]
\begin{center}
%\vspace{1cm}
        \includegraphics[width = 0.65\textwidth]{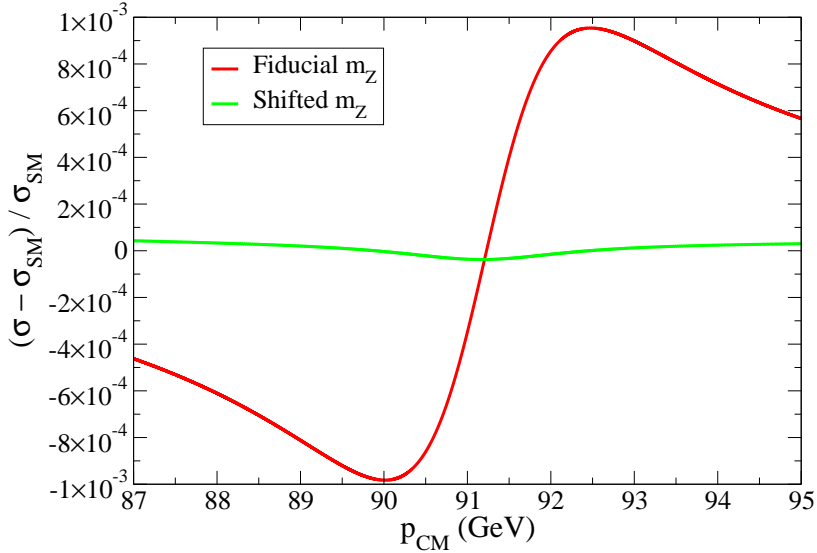}
\end{center}
\caption{Fractional correction in $\sigma(e^+e^-\to \mu^+\mu^-)$
relative to the SM for $\epsilon_0 = 10^{-2}$, $k/M_* = 0.1$, and 
$R = \pi/\gev$.  We also show the fractional correction after
shifting the input value of $m_Z$ as discussed in the text.}
\label{zpole}
\end{figure}

  We show the effect of the hidden bulk vector on the $e^+e^-\to \mu^+\mu^-$
cross-section around the $Z$ pole in Fig.~\ref{zpole}, 
where we have used $\tilde{m}_Z = 91.187\,\gev$ as a fiducial input value.
The peak of the modified cross-section is shifted away from the SM
peak by $0.0012\,\gev$, about half the current uncertainty  
in the $Z$ mass of $\Delta m_Z = 0.0023\,\gev$~\cite{Amsler:2008zzb}.
We find that the change in the lineshape can be almost completely
eliminated by changing the 
value of $m_Z$ by an amount equal to 
the shift in the peak location when the fiducial  
value is used, as we also show in Fig.~\ref{zpole}.   
After changing the  
value of $m_Z$ in this way, the $Z$ lineshape
(as well as the shifted $Z$ mass) is consistent with cross-section
measurements around the $Z$-pole at LEP~\cite{Abbiendi:2000hu}.  
We use the shifted $Z$ mass 
as an input 
in computing other electroweak observables.

%---------------------------------------------------------
%---------------------------------------------------------  

    Having fixed the input parameters $G_F$, $\alpha_{em}(Q^2\to 0)$,
and $m_Z$, we find that the modifications to other $Z$-pole observables
are completely negligible relative to current experimental
uncertainties for $k/M_*= 0.1$, $|\epsilon_0| = 10^{-2}$, and $R = \pi/\gev$.  
For all the forward-backward and polarization asymmetries, 
we obtain shifts less than $2\times 10^{-4}$ at the $Z$ pole, 
well below the current $10^{-3}$ level of sensitivity~\cite{Alcaraz:2006mx}.  
This is not entirely unexpected since the dominant mixing of the hidden bulk 
vector near the $Z$ pole is with the $Z$~\cite{McDonald:2010iq}.  As a
result, the relative 
couplings of the hidden vector to SM fermions are nearly identical to 
those of the $Z$ and the effect on the asymmetries is tiny.  
Changes to electroweak observables away from the $Z$ pole,
such as $\sigma(e^+e^-\to hadrons)$~\cite{Janot:2004cy} 
and Bhabha scattering~\cite{Karlen:2001hw},
are 
also much smaller than the precision to which they
have been measured for these model parameters~\cite{Hook:2010tw}. 

  Our results are consistent with 
Refs.~\cite{Hook:2010tw,Chang:2006fp,Feldman:2007wj}
which considered the precision electroweak bounds on a single
4D Abelian hidden vector.  For smaller values of $\epsilon_0$, the effects
on electroweak observables will be even less.  Stronger bounds
can arise when the IR scale is larger, since now the very narrow
$n=0,\,1$ KK modes may be individually resolvable.  The precise
constraints in this case can be read off from the analysis
of Ref.~\cite{Hook:2010tw}.

\subsection{Rare $Z$ Decays}

  Another source of constraints, as well as potential new signals, 
are $Z$ decays into light hidden-sector states.  This can occur
when a $Z$ mixes into a bulk vector $X$ that escapes into the extra
dimension.  The bulk vector will subsequently shower into further
vector and graviton modes.  Showering will cease when the invariant
momentum scale approaches the hidden IR scale.  At this point we can
view the products of the hidden shower as a large multiplicity
of light $n=0,\,1$ KK modes.  
In turn, these will
decay back to the SM.  The final state of such a decay mode will therefore
consist of a large multiplicity of SM 
mesons and leptons~\cite{Csaki:2008dt,Strassler:2008bv}.

%---------------------------------------------------------
\begin{figure}[ttt]
\begin{center}
%\vspace{1cm}
        \includegraphics[width = 0.75\textwidth]{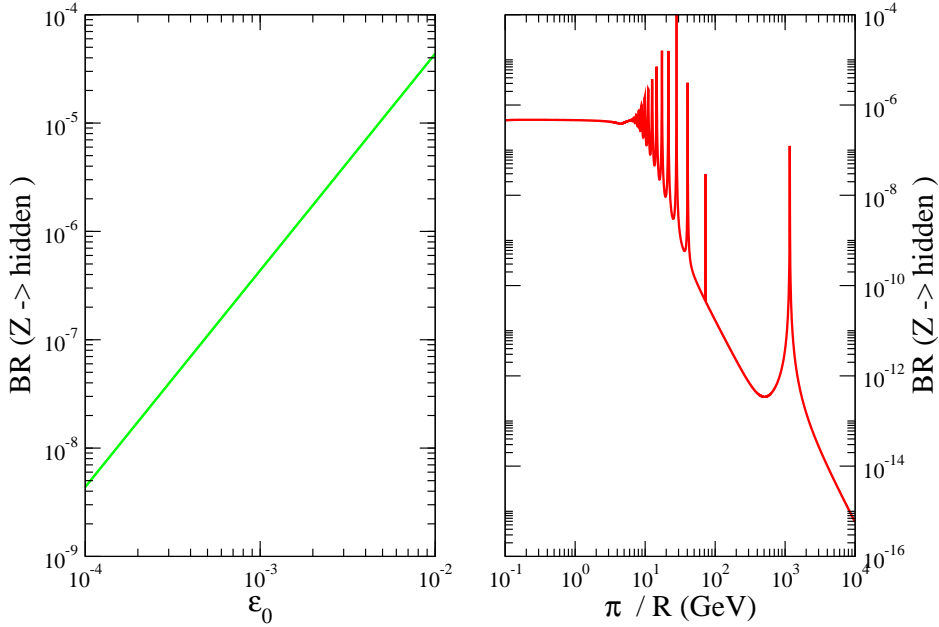}
\end{center}
\caption{Total branching fraction for $Z \to hidden$ as a function of 
$\epsilon_0$ (left) and $\pi/R$ (right).  Both panels have $k/M_* =0.1$,
while the left panel also has $\pi/R = 1\,\gev$ and the right panel
has $\epsilon_0 = 10^{-3}$.
}
\label{brzh}
\end{figure}
%--------------------------------------------------------- 

  We can compute the total \emph{inclusive} rate of such processes by 
applying a unitarity cut to the matched bulk propagator of 
Eq.~\eqref{prop_emp} connected to incoming and outgoing $Z$ boson legs.  
This yields
\bea
BR(Z\to hidden) &=& -\frac{m_Z^4}{m_Z\Gamma_Z}
\frac{\epsilon_*^2}{M_*}s_W^2\,Im(\Delta_p^{UV})\nonumber
\\
&\simeq& 
\sum_n\frac{\epsilon_n^2s_W^2m_Z^4}{(m_Z^2-m_n^2)^2+m_Z^2\tilde{\Gamma}_n^2}
\lrf{\tilde{\Gamma}_n}{\Gamma_Z}.\label{hid_z_decay_BR}
%\nnmb
\eea
In Fig.~\ref{brzh} we show the resulting branching fraction as a function
of the zero-mode mixing $\epsilon_0$ for $\pi/R = 1\,\gev$ in the left
panel, as well as the branching as a function of the KK mass splitting
$\pi/R$ for $\epsilon_0=10^{-3}$ in the right panel.  Both panels
also have $(k/M_*) = 0.1$, but the result is insensitive to the precise
value for $(M_*/k)R^{-1} \ll m_Z$. 

  For $\epsilon_0 \leq 10^{-2}$ the branching fraction is too small 
to modify the total $Z$ width in an appreciable way.  Even so, 
branching fractions greater than a few times $10^{-6}$, corresponding
to $\epsilon_0 \gtrsim 10^{-3}$, could be visible as rare events in the 
LEP~I data set~\cite{Acton:1993kp}.  Note that this is potentially much
more sensitive than indirect bounds from $e^+e^-\to f\bar{f}$ processes.
However, the high-multiplicity final states predicted by this scenario 
were not explicitly searched for, and thus we do not quote a precise limit 
on $\epsilon_0$.  When the KK mode spacing is significantly larger than 
a $\gev$, mixing of individual KK modes with the $Z$ can be 
resonantly enhanced~\cite{Hook:2010tw}.

  Observe that the branching fraction in Fig.~\ref{brzh} becomes
independent of $R$ for $R^{-1}\lesssim 10$~GeV. This feature
is readily understood; for $m_Z\gg (M_*/k)R^{-1}$ 
the bulk propagator does not probe the deep IR and is well
approximated by the $R$-independent RS2 form. Consequently a simple
expression for the hidden $Z$-width can be obtained. Using
Eq.~\eqref{rs2prop} in Eq.~\eqref{hid_z_decay_BR} 
the branching fraction in this regime can be approximated
as
\bea
BR(Z\to hidden) &\simeq& 
%5%
%\frac{\pi}{2}s_W^2\epsilon_*^2\lrf{k}{M_*}\frac{1}{[\log(2k/m_Z)-\gamma]^2}
%\lrf{m_Z}{\Gamma_Z}\\
%&\simeq& 
\frac{\pi}{2}\frac{k}{M_*}\,
\frac{s_W^2\epsilon_*^2}{[\log(2k/m_Z)-\gamma]^2}
%5% - the expression is now correct
%6%\left[\frac{\log(2k/m_0)-\gamma}{\log(2k/m_Z)-\gamma}\right]
%5%
\lrf{m_Z}{\Gamma_Z}\ \nonumber\\
&\simeq &
\frac{\pi}{2}\,
\frac{s_W^2\epsilon_0^2}{[\log(2k/m_Z)-\gamma]}\lrf{m_Z}{\Gamma_Z}\
\sim\  0.4 \ \epsilon_0^2\ .
%\nnmb
 %
%\frac{\epsilon_0^2s_W^2}{36}\frac{m_Z}{ \Gamma_Z}\simeq \frac{10^{-6}}{6},
\label{simple_hid_z_decay_BR}
\eea
We note that among
the parameters
describing the hidden sector, this result is sensitive to
$\epsilon_0$ alone (up to a very weak logarithmic $k$ dependence).  
Since this parameter also controls the direct production of the lightest 
vector mode in low-energy experiments, the model predicts an important 
correlation between these different observables.

%%%%%%%%%%%%%%%%%%%%%%%%%%%%%%%%%%%%%%%%%%%%%%%%%%%%%5

\section{Low-Energy Measurements\label{sec:lowe}}

  Low-energy experiments with a very high precision or luminosity
can in many cases provide a much more sensitive test of the present
scenario than higher-energy collider experiments.  Specific examples
include measurements of the anomalous magnetic moment of the muon,
fixed-target and beam dump experiments, and meson factories.
In this section we investigate the power of such lower-energy
probes to test a light warped hidden sector.

\subsection{Sub-$\mathbf{GeV}$ Constraints and Searches}

  Experimental constraints on a single Abelian hidden vector
have been discussed extensively in Refs.~\cite{Batell:2009yf,
Bjorken:2009mm,Pospelov:2008zw}.
For vector masses below a GeV the strongest model-independent
bounds come from the anomalous $e$ and $\mu$ magnetic moments for larger 
kinetic mixing $\epsilon$~\cite{Pospelov:2008zw}, 
together with null beam dump searches for smaller 
kinetic mixing~\cite{Bjorken:2009mm}.  Vector masses as small as
$10\,\mev$ are allowed for $\epsilon \lesssim 10^{-3}$.  Much lighter
vectors can also be consistent with existing bounds for extremely
small values of $\epsilon$~\cite{Ahlers:2008qc}, but we do not consider 
this possibility here.

  In the present scenario, the single vector is replaced by a tower
of KK resonances.  Even so, with the exception of meson factories, 
the corresponding low-energy experimental bounds on the tower are 
almost completely dominated by the lightest zero mode.  
We find that these bounds are nearly identical to the case of 
a single Abelian hidden vector with the same mass and kinetic 
mixing as the zero mode.  This occurs for two reasons.  First, the constraints
due to leptonic magnetic moments and beam dump experiments generally
become weaker as the vector mass increases.  Second, and usually 
more importantly, the zero mode has a significantly larger kinetic 
mixing coupling to the SM than the heavier KK modes, as can be seen 
by comparing Eqs.~\eqref{epsilonz} and \eqref{epsilonn}.

  Summarizing these limits, for $m_0$ below $10\,\gev$ the corresponding bounds can be obtained by applying 
the constraints of Refs.~\cite{Bjorken:2009mm,Pospelov:2008zw} to the 
zero mode vector. For such light masses $\epsilon_0 \lesssim 3\times 10^{-3}$ is needed
to satisfy the constraints from leptonic magnetic moments~\cite{Pospelov:2008zw}
and searches for exotic events within the BABAR $\Upsilon(3S)$ 
data set~\cite{Bjorken:2009mm}.  Masses below about $0.5\,\gev$
are further constrained by beam dump searches and supernova cooling,
leading to bounds on smaller values of $\epsilon_0$.  Together, there remains
a pocket of allowed $m_0$-$\epsilon_0$ values in the range of
$m_0 = 10\,\mev$-$10\,\gev$ as exhibited in Ref.~\cite{Bjorken:2009mm}.
Masses below $m_0 \simeq 10\,\mev$ are only consistent with very small
values of the kinetic mixing, 
below $\epsilon_0 \lesssim 5\times 10^{-8}$~\cite{Bjorken:2009mm,Ahlers:2008qc}.
Once these low-energy constraints are satisfied (for $m_0 \lesssim 10\,\gev$)
the constraints from precision electroweak data are are also met.

  Current and planned fixed-target experiments will provide even more
stringent bounds on light hidden vectors in the near future.
Hidden vector production in such experiments occurs dominantly in the 
forward direction as it is enhanced by a collinear singularity cut off
by the vector mass. Thus the lightest KK modes will be produced 
most abundantly.  The fixed-target search techniques proposed in 
Refs.~\cite{Essig:2009nc,Reece:2009un,Bjorken:2009mm}
are suited for hidden vectors that decay directly to a pair of SM fermions.
These searches can therefore also be sensitive to the $n=0,1$ KK modes
of a bulk hidden vector.  Modified search techniques will likely be
needed to detect the higher KK modes, which become progressively
broader and decay to complicated multi-body final states.

%%%%%%%%%%%%%%%%%%%%%%%%%%%%%%%%%%%%%%%%%%%%%%%%%%%%%%%55
%---------------------------------------------------------
\begin{figure}[ttt]
\begin{center}
%\vspace{1cm}
        \includegraphics[width = 0.35\textwidth]{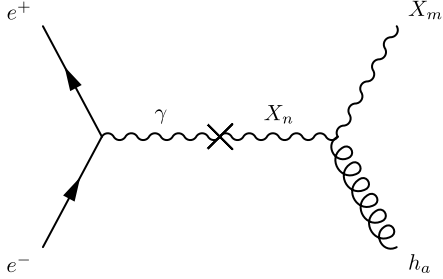}
\end{center}
\caption{Hidden sector production in the $s$-channel. This only requires one
  kinetic mixing insertion (represented by the cross) and therefore
  scales as $\epsilon^2$, improving the prospects for
  production. Resonant enhancement can occur when $\sqrt{s}\simeq m_n$ for
one of the 
small-$n$ modes.}
\label{fig:s}
\end{figure}
%---------------------------------------------------------

\subsection{New Searches at Meson Factories\label{sec:meson_factory}}  

  Meson factories like DA$\Phi$NE, BABAR
and Belle provide some of the most promising means by which to probe 
the warped hidden sectors we consider. The %precise 
specific
signals depend on the value of the hidden IR scale relative to the 
center-of-mass (CoM) energy $\sqrt{s}$ of the given experiment. 
If $\sqrt{s} \simeq m_n$ for one of the  narrow small-$n$ 
KK modes,
hidden vector production can be resonantly enhanced and specific
exclusive signatures can be searched for.  
However at higher CoM energies $\sqrt{s} \gtrsim (M_*/k)R^{-1}$,
well above the mass of the narrow lighter KK modes, 
one instead probes the continuum part of the spectrum and 
there is no resonant
enhancement. In this regime the typical signal consists of a large
multiplicity of soft SM particles. 

  The approaches discussed for discovering 
a single Abelian hidden vector in 
Refs.~\cite{Batell:2009yf,Essig:2009nc,Reece:2009un,Fayet:2007ua} 
also apply to the 
narrow $n=0,\,1$ KK modes in the present scenario when they decay primarily
 to the SM.  
These searches focus on single vector production via
$e^+e^-\rightarrow \gamma X_n$. 
The hidden vector
decays to a pair of SM leptons or pions, giving signals like
$\ell^+\ell^-\gamma$. The tiny vector width 
leads to a distinctive dilepton invariant mass peak that can
potentially be distinguished from the smooth SM background.
A detailed search for this signal within the BABAR and Belle 
$\Upsilon(3s)$ and $\Upsilon(4s)$ datasets could potentially probe hidden 
vectors with kinetic mixing as low as 
$10^{-3}$~~\cite{Essig:2009nc,Reece:2009un,Bjorken:2009mm}.
These searches will be most sensitive to the $n=0$ mode
as the single vector production rate scales like $\epsilon_n^2
\simeq \epsilon_0^2/(36n^2)$, (see~(\ref{epsilonz}) and (\ref{epsilonn})). 
More complicated multi-lepton or pion final states can also 
arise for the $n=0,1$ modes if there is a light hidden 
Higgs~\cite{Batell:2009yf,Gopalakrishna:2008dv,Fayet:2007ua}.

  The same production channel can also be used to probe narrow $n>1$ 
vector modes that decay mainly into the hidden sector before cascading
back to the SM. For example, production of an $n=2$ vector mode would give
\bea
e^+e^-\rightarrow \gamma \ X_2\rightarrow\gamma\  X_0\ h_1\ .
\eea
The KK graviton decays via $ h_1 \to 2 X_0$ so the final state 
consists of $\gamma+6\ell$ (for leptonic $n=0$ KK decays). 
Similar statements hold for single vector 
production in Kaon decays~\cite{Pospelov:2008zw,Reece:2009un}.  
Again, even higher multiplicities in the final state can arise 
if there is an explicit light IR 
Higgs.

  Hidden sector production can be resonantly enhanced when $\sqrt{s} \sim m_n$.
For two reasons, this enhancement is not significant for hidden vectors 
that decay entirely to the SM, such as the $n=0,\,1$ KK vectors.
First, the relevant process in this case is $e^+e^-\to X \to f\bar{f}$,
and its rate is proportional to $\epsilon^4$.  Second, the resonance
is typically much narrower than the beam energy spread and gets
strongly smeared out.  In the present scenario, however, the KK gravitons 
in the spectrum permit more efficient $s$-channel production processes,
such as that of Fig.~\ref{fig:s}, %$e^+e^-\to X_n \to X_mh_a$ 
which only require one kinetic
mixing insertion and thus go like $\epsilon^2$.  The graviton decay modes
of a hidden vector also broaden its resonance.  This is analogous
to multi-vector production in non-Abelian hidden 
gauge sectors~\cite{Essig:2009nc},
as one might 
expect from gauge-gravity duality.

%---------------------------------------------------------
\begin{figure}[ttt]
\begin{center}
%\vspace{1cm}
        \includegraphics[width = 0.7\textwidth]{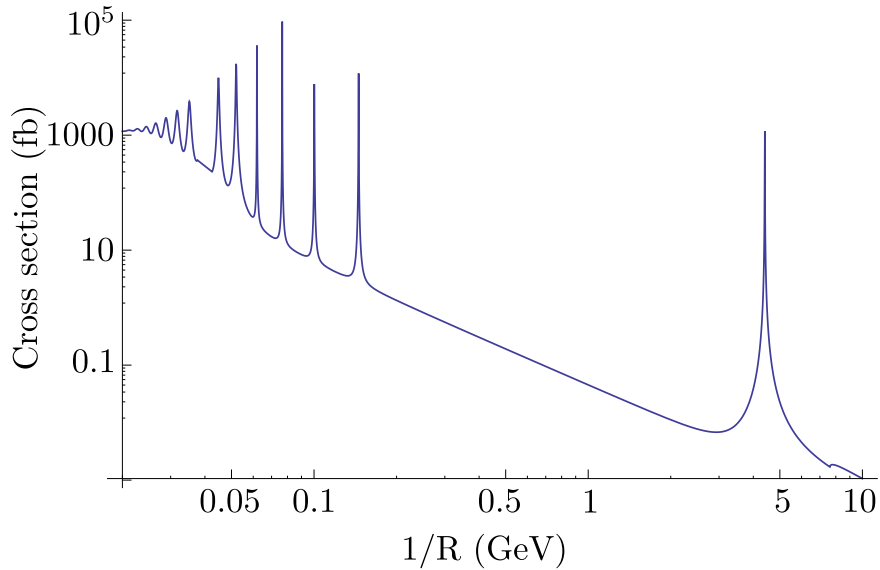}
\end{center}
\caption{Inclusive cross section for hidden sector production at the
  center-of-mass energy of DA$\Phi$NE ($\sqrt{s}=1.02$~GeV) as a
  function of the hidden IR scale $R^{-1}$. The plot is for the
  Higgsless case with the kinetic
  mixing parameter for the zero mode fixed at
  $\epsilon_0=10^{-3}$.}
\label{fig:hid_prod_102Gev}
\end{figure}
%---------------------------------------------------------

\begin{figure}[ttt]
\begin{center}
        \includegraphics[width = 0.75\textwidth]{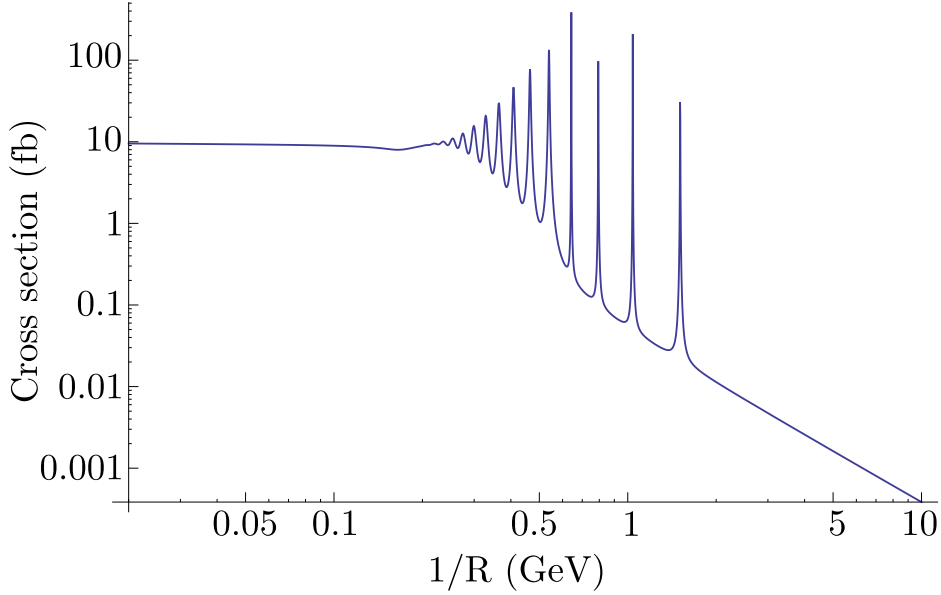}
\end{center}
\caption{Same as Figure~\ref{fig:hid_prod_102Gev} but for
  the center-of-mass energy of the B-factories ($\sqrt{s}=10.58$~GeV).}
\label{fig:hid_prod_1058Gev}
\end{figure}

  We estimate the prospects for observing resonant 
production by calculating the inclusive hidden-sector cross section
at meson factories.  At the relatively low energies ($1$--$10\,\gev$)
of these machines
we may neglect diagrams involving the $Z$ boson and consider 
the insertion of a kinetic mixing operator between a hidden 
vector $X_n$ and the SM photon. 
Applying a unitarity cut gives the inclusive hidden sector cross
section as 
\bea
\sigma (e^+e^-\rightarrow Hidden)
%&=&
%-\frac{e^2c_W^2\epsilon_*^2}{s\,M_*}Im(\Delta^{UV}_{\sqrt{s}})
%~~~~~\,,\quad (s\ll m_Z^2)\\
&\simeq&
\sum_n \frac{e^2c_W^2\epsilon_n^2\,\tilde{\Gamma}_n\,\sqrt{s}}
{(s-m_n^2)^2+s\,\tilde{\Gamma}^2_n}\,,\quad (s\ll m_Z^2)
\label{hidcs}
\eea
where $s$ is the usual Mandelstam 
variable.
We plot this inclusive cross section as a function of $1/R$ 
at the CoM energy of DA$\Phi$NE ($\sqrt{s}=1.02$~GeV) in 
Fig.~\ref{fig:hid_prod_102Gev} and for the B-factories 
($\sqrt{s}=10.58$~GeV) in Fig.~\ref{fig:hid_prod_1058Gev}. 
In both figures we fix $\epsilon_0=10^{-3}$.

 %---------------------------------------------------------
%--------------------------------------------------------- 

  The peaks in Figs.~\ref{fig:hid_prod_102Gev} and \ref{fig:hid_prod_1058Gev}
occur at $s$-channel KK resonances, 
 when $1/R$ is such that 
$\sqrt{s}\simeq m_n$ for one of the narrow, small $n$ KK modes.  For
example, the peak at $R^{-1} \simeq 5$~GeV 
in Fig.~\ref{fig:hid_prod_102Gev} occurs because DA$\Phi$NE would be 
sitting right on the zero mode mass of $m_0\simeq 1$~GeV.  
In practice, this very narrow peak will be significantly smeared
out by the spread in beam energy; about $0.2\,\mev$
at DA$\Phi$NE~\cite{Bossi:2008aa} and a few MeV at the 
B-factories~\cite{Aubert:2001tu}.  As a result, it is unlikely
to be visible as a direct $s$-channel 
resonance~\cite{Reece:2009un,Pospelov:2008zw}.
The same also applies to the $n=1$ mode when it decays entirely 
to the SM via kinetic mixing.
Higher vector KK modes have decays to the hidden sector, making them
much broader and relatively insensitive to the energy spread
of the meson factory beams.
As $1/R$ decreases, the peaks become more and more closely-spaced
and less sharp.  This corresponds to resonances occurring at ever 
larger KK mode numbers, which eventually begin to overlap 
with each other.  

  For  $\sqrt{s}\gg M_*/kR $ the inclusive
cross-section becomes
roughly independent of $1/R$, as can be seen in
Fig.~\ref{fig:hid_prod_1058Gev}  (similar behaviour would
be seen in Fig.~\ref{fig:hid_prod_102Gev} at smaller values of $1/R$).
This reflects the IR insensitivity 
of the theory for large injection energies on the UV brane, 
as discussed in Section~\ref{sec:match}.  In this region 
the inclusive cross section can be written in a simple form by
using the RS2 propagator of Eq.~\eqref{rs2prop} in Eq.~\eqref{hidcs}: 
\bea
\sigma (e^+e^-\rightarrow Hidden) 
&\simeq& \frac{\pi}{2}\frac{k}{M_*}\frac{ e^2 
c_W^2\epsilon_*^2}{[\log(2k/\sqrt{s})-\gamma]^2}\ 
\frac{1}{s}\nonumber\\
&\simeq&\frac{\pi}{2}\frac{ e^2 
c_W^2\epsilon_0^2}{[\log(2k/\sqrt{s})-\gamma]}\ 
\frac{1}{s}
\label{simple_hidd_prod_b_factories} .
\nonumber
\eea
We note that, up to a mild logarithmic sensitivity to $k$, this depends
on the single unknown parameter $\epsilon_0$ and can be written as:
\bea
\sigma (e^+e^-\rightarrow Hidden) 
&\simeq&
\left(\frac{\epsilon_0}{10^{-3}}\right)^2\times\left(\frac{10.58~\mathrm{GeV}}{\sqrt{s}}\right)^2\times 10\ fb\ .
\nonumber
\eea
For $\epsilon_0 = 10^{-3}$ 
one obtains an asymptotic small-$1/R$ inclusive cross section of
$\sigma (e^+e^-\rightarrow Hidden)\simeq 10~fb$ for the $B$-factories,
as seen in Fig.~\ref{fig:hid_prod_1058Gev}.

  Once created, hidden states will cascade down to the lightest 
vector modes, $n=0,\,1$.  Provided the kinetic
mixing is not too small, $\epsilon_{0,1} \gtrsim 10^{-4}$,  
these lightest hidden modes will decay relatively 
promptly to the SM.  
Thus a typical final state will consist of a number of SM fields 
with pairwise invariant masses equal to one of the KK vector masses.
A light hidden Higgs or smaller values of $\epsilon_{0,1}$ can
also give rise to displaced vertices.

%---------------------------------------------------------
\begin{figure}[ttt]
\begin{center}
        \includegraphics[width = 0.4\textwidth]{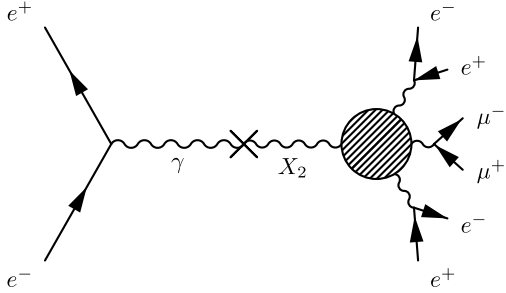}
\end{center}
\caption{Production of six-lepton final states in the
  $s$-channel. Hidden sector production occurs via $e^+e^-\rightarrow
  h_1X_0$ and the KK graviton promptly
  decays to $2X_0$ (collectively represented by the blob).
  The 3$X_0$'s in turn decay to six
  light SM fields. Production is only suppressed by $\epsilon^2$ and
  can be resonantly enhanced for
  $\sqrt{s}\simeq m_2$. Similar processes lead to eight-lepton final states.}
\label{fig:6fermi}
\end{figure}
%---------------------------------------------------------

  The optimal detection strategy at meson factories 
depends on the quantity $R\sqrt{s}$.  For $R\sqrt{s} \lesssim 1$, 
only the lightest modes will be produced in association with a photon,
with $n=0$ modes dominating.  The resulting final state will consist
of a pair of SM particles reconstructing the KK vector mass.
For larger $R\sqrt{s}$
 there can be resonant production of hidden
vectors in the $s$-channel.   The final states in this case will
consist of a number of SM fields with pairwise invariant masses equal 
to the mass of the mode $n=0$ or $n=1$. The precise number of
final-state SM
particles increases with $R\sqrt{s}$, but as the multiplicity increases
 so too will the combinatoric problem of reconstructing pairwise
invariant mass peaks.  On the other hand, a high multiplicity of 
charged pions and leptons should make these events very distinctive.
Scanning the inclusive multiparticle cross section in energy may allow 
one to probe heavier KK vector resonances that are not overly broad.

  More stringent bounds on warped
  Abelian hidden sectors (or possible discovery signals)
  could potentially be obtained 
 with existing data. 
 New searches for narrow resonances in
 four-lepton final states have recently been undertaken
 by the BABAR Collaboration~\cite{:2009pw}. Their analysis demands
 that the four leptons reconstruct to (nearly) the full beam energy,
 resulting in very strong bounds of 
 $\sigma(e^+e^-\rightarrow 2\ell2\ell')<(25-60)~ab$ for hidden vector 
 masses in the range $[0.24,5.3]$~GeV. This final state
 can result from the $s$-channel process $e^+e^-\rightarrow
 W'\ W'\rightarrow 2\ell\ 2\ell'$ in theories with a
 non-Abelian hidden sector, but does not occur through the $s$-channel in
 our warped model.  Four-lepton final states
 can occur in the present model via light vector production in the $t$-channel,
 $e^+e^-\rightarrow 2X_{0,1}$, with the final state vectors decaying 
 to lepton pairs. However, due to the $\epsilon^4$ suppression the
 BABAR bound is not severe. 
 
On the other hand, improved analysis of six-lepton final states like 
\bea
 e^+e^-\rightarrow2\mu 4e,\ 4\mu 2e,\  6\mu,
 \eea
could improve the bounds on the warped model. These occur 
 via hidden sector production in the $s$-channel with only
 $\epsilon^2$ suppression (see
 Fig.~\ref{fig:6fermi}), permitting more efficient production. The same
 is true of eight-lepton final states
 from processes
 like $e^+e^-\rightarrow h_2X_0\rightarrow 4X_0$. Given the strong
 bounds from the four-lepton BABAR analysis~\cite{:2009pw} relative to
 the typical cross sections shown in Fig.~\ref{fig:hid_prod_1058Gev},
 similar experimental studies for $N>4$ final-state leptons 
could greatly improve the prospects for detection. 
  One hopes that such analysis
 will be forthcoming. Note that with a total integrated
 luminosity of $\sim1~ab$, a production  cross section of order a
 $fb$ corresponds to 
a few hundred to a few thousand hidden sector production events.

  Together these features suggest an interesting 
multi-faceted approach to experimentally studying the model 
for $m_ZR\gg1$. One
may be able to probe individual resonances directly at low-energy colliders
and fixed target experiments operating at order GeV energies, and
extract information on the kinetic mixing parameter $\epsilon_0$. 
Via Eq.~(\ref{simple_hidd_prod_b_factories}) one could then predict 
the expected inclusive cross section for hidden sector production at the
$B$-factories. Furthermore these two probes can be combined with an
expected signal from hidden $Z$ decays as given in
Eq.~(\ref{simple_hid_z_decay_BR}). These correlations provide important
means by which to discriminate the present framework from alternative
light hidden sectors.

%%%%%%%%%%%%%%%%%%%%%%%%%%%%%%%%%%%%%%%%%%%%%%%%%%%%%%%%%%%%%%%%%%%%%%%5

\section{Conclusion\label{sec:concl}}

In this work we have investigated the detailed phenomenology of an
Abelian warped hidden sector, focusing on hidden symmetry
breaking scales much less than the weak scale. Such light hidden sectors may
be of interest in
connection with recent models of dark matter, but more generally comprise an
interesting scenario for beyond-the-SM 
physics. Despite the relative
simplicity of our model the low-energy
phenomenology was seen to be quite rich. The main new feature
of our construct, relative to other works with light hidden sectors,
is the existence of a tower
of hidden KK vectors that kinetically mix with SM hypercharge.

We have considered the decay properties of the hidden sector fields in some
detail, and developed a useful approach for calculating processes
initiated on the UV brane with large injection momentum relative to
the hidden IR scale. Using these results, we have considered the 
detailed bounds on the model from precision electroweak observables 
and low-energy experiments, and found that viable parameter sets 
permitting significant numbers of hidden sector
production events exist. 

  For a hidden IR scale in the range 10~MeV to 10~GeV, we find that 
the constraints on a tower of hidden vectors are nearly identical to
those on a single hidden vector whose mass and kinetic mixing
equals that of the zero mode. This is primarily due to the properties 
of the KK vectors: the higher modes are heavier and,
more importantly, their kinetic mixing strength decreases 
(relative to the zero mode).
%Low-energy constraints, which are the most severe, are therefore
%dominated by the zero mode. 
The most stringent bounds are obtained by
applying  existing constraints from
beam dumps, $\Upsilon$ decays, and anomalous lepton magnetic 
moments directly to the zero mode. The precise bounds can be read off
the figures in Refs.~\cite{Bjorken:2009mm,Pospelov:2008zw}, and 
generally require $\epsilon_0\lesssim 3\times 10^{-3}$ for
$m_0\sim\mathcal{O}(\mathrm{GeV})$. Once the zero mode is made to
satisfy these low-energy constraints, compatibility with precision
electroweak data is ensured. 

  Relative to models with a single hidden vector there are, however,  
some important differences that can have implications for future
searches. The hidden sector contains multiple light
vectors that can potentially be probed as individual resonances at low-energy
experiments like the $B$-factories and fixed-target experiments for
values of the kinetic mixing parameter that are consistent with 
existing direct search constraints.
The model also predicts hidden sector decays of the 
$Z$-boson with rates essentially dependent on the single parameter 
$\epsilon_0$ for the interesting range $m_ZR\gg1$. 
As this parameter controls the production and decay rate
of the lightest vectors, important correlations exist between the low-energy
signals and the hidden $Z$ width. Future experimental 
studies of six- and
eight-lepton final states could also improve the bounds on the model, or
potentially discover evidence for a hidden warped extra dimension.

%%%%%%%%%%%%%%%%%%%%%%%%%%%%%%%%%%%%%%%%%%%%%%%%%%%%%%%%%%%%

\section*{Acknowledgements}

  We thank Kaustubh Agashe, Brian Batell, 
Robert McPherson,
John Ng, Lisa Randall, 
Steve Robertson,
Matt Schwartz, Jessie Shelton, Andy Spray, Matt Strassler, 
Jay Wacker, and Lian-Tao Wang
for helpful discussions.  KM thanks the theory group at the University of
Melbourne for hospitality while parts of this work were completed. 
DM thanks the Aspen Center for Physics for their hospitality.
This work was supported by NSERC.

%%%%%%%%%%%%%%%%%%%%%%%%%%%%%%%%%%%%%%%%%%%%%%%%%%%%%%%%%%%%%%%%%%%%%%

\section*{Appendix}
\appendix

%%%%%%%%%%%%%%%%%%%%%%%%%%%%%%%%%%%%%%%%%%%%%%%%%%%%%%%%%%%%%

\section{Hidden Sector Decays}
In this appendix we present the decay widths for KK vectors and KK gravitons into the hidden sector. 
\subsection{Vector Decays\label{app:vector_decay}}
  To determine the coupling between KK vectors and gravitons one expands the metric as
\begin{eqnarray}
G_{\mu\nu} \rightarrow (kz)^{-2}\left[\eta_{\mu\nu}
+\frac{2}{M_*^{3/2}}{h}_{\mu\nu}(x,z)\right],
\end{eqnarray}
where we work in the gauge $\partial^{\mu}h_{\mu\nu} = 
0 = h^{\mu}_{\mu}$. KK expanding the graviton fluctuation as
\begin{eqnarray}
h_{\mu\nu}(x,z)=\sum_{a}h^{(a)}_{\mu\nu}(x)f_{h}^{(a)}(z),\label{h_kk}
\end{eqnarray}
 the effective 4D Lagrangian contains the following coupling between the KK gravitons and vectors:
\begin{eqnarray}
\mathcal{L}_{eff} \supset \frac{k}{M_{Pl}}\sum_{a,m,n}
\eta^{\rho\nu}\eta^{\sigma\beta}\;h_{\rho\sigma}^{(a)}\left(
{\zeta_{a,mn}}\eta^{\mu\alpha}X^m_{\mu\nu}X^n_{\alpha\beta}
-\xi_{a,mn}X^m_{\nu}X^n_{\beta}\right).
\label{gravicoup}
\end{eqnarray}
Here the factors $\zeta_{a,mn}$ and $\xi_{a,mn}$ encode the wavefunction overlap along the extra dimension,
\begin{eqnarray}
\zeta_{a,mn} &=& \frac{1}{k^{3/2}}\int\!\frac{dz}{(kz)}\,
f^{(a)}_{h}\,f^{(m)}_X\,f^{(n)}_X\ \ ,\\
\xi_{a,mn} &=& \frac{1}{k^{3/2}}\int\!\frac{dz}{(kz)}\,
f^{(a)}_{h}\,\partial_zf^{(m)}_X\,\partial_zf^{(n)}_X\ \ ,
\end{eqnarray}
and $f^{(m)}_X,f^{(n)}_X$ are the KK vector profiles. The vertex derived from Eq.~(\ref{gravicoup}) is
\begin{eqnarray}
& &h_{\rho \sigma}^{(a)}(p)X^m_{\nu}(k_1)X^n_\beta(k_2):\nonumber \\
& &-2i\frac{k}{M_{Pl}}\left\{\eta^{\rho\nu}\eta^{\sigma\beta} [\xi_{a,mn}+k_1\cdot k_2 \zeta_{a,mn}]+\zeta_{a,mn}[ \eta^{\nu\beta}k_1^\rho k_2^\sigma -\eta^{\rho\nu}k_1^\beta k_2^\sigma -\eta^{\sigma \beta}k_1^\rho k_2^\nu ]\right\},
\end{eqnarray}
where the graviton momentum $p$ is defined as incoming and the vector
momenta $k_{1,2}$ are outgoing. This vertex can be used to calculate the decay width for $X_n\rightarrow X_m h_{a}$. We find the width to be
\begin{eqnarray}
& &\Gamma (X_n\rightarrow X_mh_{a})\nonumber\\
&=&\frac{1}{144\pi}\frac{k^2}{M_{Pl}^2}\frac{m_n^7}{m_a^4}\times[1-(r_m+r_a)^2]^{1/2}[1-(r_m-r_a)^2]^{1/2}\nonumber\\
& &\times \left\{2\zeta_{a,mn}^2 G^\zeta_{a,mn}+40\zeta_{a,mn}\left[\frac{\xi_{a,mn}}{m_mm_n}\right]G^{\zeta \xi}_{a,mn}+\left[\frac{\xi_{a,mn}}{m_mm_n}\right]^2G^{\xi}_{a,mn}\right\},\label{hidden_kkx_decay}
\end{eqnarray}
where we write the mass ratios as $r_{m,a}=m_{m,a}/m_n$ and define the following set of constants
\begin{eqnarray}
G^\zeta_{a,mn}&=&1+2(r_a^2-2r_m^2)+(6r_m^4-r_m^2r_a^2+r_a^4)-(4r_m^6+r_a^2r_m^4-34r_a^4r_m^2+9r_a^6)\nonumber\\
& &+\ (r_m^8+r_a^2r_m^6+r_a^4r_m^4-9r_a^6r_m^2+6r_a^8),\nonumber\\
G^{\zeta \xi}_{a,mn}&=&r_mr_a^2\left\{1+(r_a^2-2r_m^2)+(r_m^4+r_m^2r_a^2-2r_a^4)\right\},\nonumber\\
G^{\xi}_{a,mn}&=&1+2(3r_a^2-2r_m^2)-2(7r_a^4+3r_a^2r_m^2-3r_m^4)+2(3r_a^6+42r_m^2r_a^4-3r_m^4r_a^2-2r_m^6)\nonumber\\
& &+\ (r_a^2-r_m^2)^2(r_m^4+8r_a^2r_m^2+r_a^4).
\end{eqnarray}

We note that the sum over graviton polarizations necessary for calculating the KK decays is~\cite{Han:1998sg}
\begin{eqnarray}
\sum_{pol.}e_{a,\rho\sigma}(p)e_{a,\mu\alpha}^*(p)&=&B_{\rho\sigma,\mu\alpha}^{(a)}(p),
\end{eqnarray}
where
\begin{eqnarray}
B_{\rho\sigma,\mu\alpha}^{(a)}(p)&=&\left(\eta_{\rho\mu}-\frac{p_\rho p_\mu}{m_a^2}\right)\left(\eta_{\sigma\alpha}-\frac{p_\sigma p_\alpha}{m_a^2}\right)+\left(\eta_{\rho\alpha}-\frac{p_\rho p_\alpha}{m_a^2}\right)\left(\eta_{\sigma\mu}-\frac{p_\sigma p_\mu}{m_a^2}\right)\nonumber\\
& &-\frac{2}{3}\left(\eta_{\rho\sigma}-\frac{p_\rho p_\sigma}{m_a^2}\right)\left(\eta_{\mu\alpha}-\frac{p_\mu p_\alpha}{m_a^2}\right).
\end{eqnarray}

%%%%%%%%%%%%%%%%%%%%%%%%%%%%%%%%%%%%%%%%%%%%%%%%%%%%%%%%%%%%%%%55

\subsection{Hidden KK Graviton Decays\label{app:decay_kk_grav}}
\begin{figure}[t]
\centering
\includegraphics[width=0.55\textwidth]{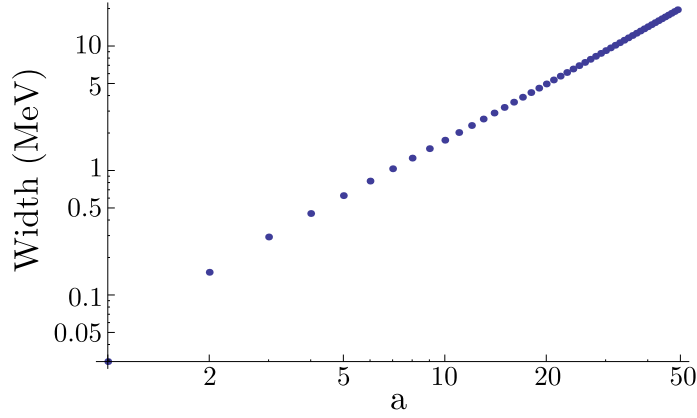} 
  \caption{Higgsless Case: Plot of $\sum_{m,n}\Gamma(h_a\rightarrow
    X_m X_n)$ Vs. KK number $a$, where the sum is over values of $m,n$
    satisfying $a>m+n$. Similar behavior is observed in the weakly
    Higgsed case.}
  \label{fig:gravitonwidth_v_daughterkk_number_nohiggs}
\end{figure}

\begin{figure}[t]
\centering
\includegraphics[width=0.55\textwidth]{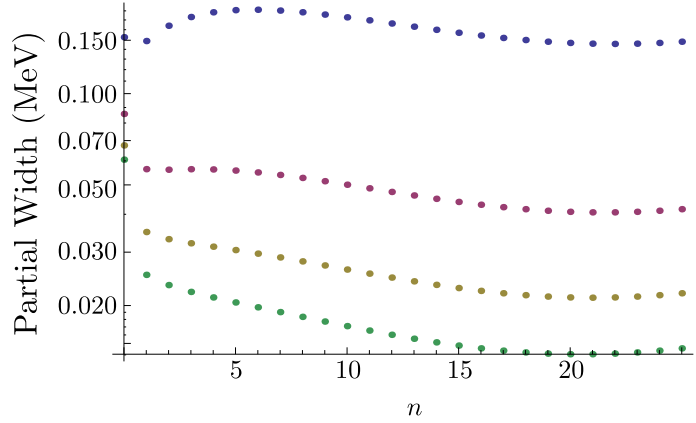} 
  \caption{Higgsless Case: Plot of $\Gamma(h_{a=45}\rightarrow X_n X_m)$ Vs. Daughter KK number $n$ for fixed values of $(m+n)$. From top to bottom the curves satisfy $(m+n)=(44,43,42,41)$.}
  \label{fig:graviton_a_45_width_v_daughterkk_number_nohiggs}
\end{figure}

The KK graviton decay $h_a\rightarrow X_m X_n$ is kinematically available  for 
\begin{eqnarray}
a\ge m+n\quad&&\quad \mathrm{Higgsed},\nonumber\\
a> m+n\quad&&\quad \mathrm{Higgsless},
\end{eqnarray}
where the  Higgsed case assumes that the Higgs VEV is less than the IR
scale so that the zero mode vector acquires a mass $m_0 \sim
(5R)^{-1}$. For $a>0$ the KK graviton masses are approximately $m_a
R\simeq (a+1/4)\pi$ and for a given value of $n>0$ there are less
available channels in the Higgsless case. The width for the decay of a
KK graviton into two vectors is
\begin{eqnarray}
& &\Gamma (h_{a}\rightarrow X_mX_n)\nonumber\\
&=&\frac{S_{mn}}{240\pi}\frac{k^2}{M_{Pl}^2}m_a^3[1-(\bar{r}_m+\bar{r}_n)^2]^{1/2}[1-(\bar{r}_m-\bar{r}_n)^2]^{1/2}\nonumber\\
& &\times \left\{12\zeta_{a,mn}^2 F^\zeta_{a,mn}+80\zeta_{a,mn}\left[\frac{\xi_{a,mn}}{m_mm_n}\right]F^{\zeta \xi}_{a,mn}+\left[\frac{\xi_{a,mn}}{m_mm_n}\right]^2F^{\xi}_{a,mn}\right\},
\end{eqnarray}
where we define the symmetry factor and mass ratios respectively as
\begin{eqnarray}
S_{mn}=[1-(1/2)\delta_{mn}]\quad,\quad \bar{r}_{m,n}=\frac{m_{m,n}}{m_a},
\end{eqnarray} 
and also define the following constants:
\begin{eqnarray}
F^\zeta_{a,mn}&=&1-\frac{3}{2}(\bar{r}_m^2+\bar{r}_n^2)+\frac{1}{6}(\bar{r}_m^4+34\bar{r}_m^2\bar{r}_n^2+\bar{r}_n^4)+\frac{1}{6}(\bar{r}_m^2-\bar{r}_n^2)^2(\bar{r}_m^2+\bar{r}_n^2)+\frac{1}{6}(\bar{r}_m^2-\bar{r}_n^2)^4\nonumber\\
F^{\zeta \xi}_{a,mn}&=&\bar{r}_m\bar{r}_n\left\{1-\frac{1}{2}(\bar{r}_m^2+\bar{r}_n^2)-\frac{1}{2}(\bar{r}_m^2-\bar{r}_n^2)^2\right\}\nonumber\\
F^{\xi}_{a,mn}&=&1+6(\bar{r}_m^2+\bar{r}_n^2)-14(\bar{r}_m^4-6\bar{r}_m^2\bar{r}_n^2+\bar{r}_n^4)+6(\bar{r}_m^2-\bar{r}_n^2)^2(\bar{r}_m^2+\bar{r}_n^2)+(\bar{r}_m^2-\bar{r}_n^2)^4.\nonumber
\end{eqnarray}
With an order GeV IR scale the KK graviton decays are prompt, as can be seen in
Fig.~\ref{fig:gravitonwidth_v_daughterkk_number_nohiggs} where we plot
the two-body decay width $\sum_{m,n}\Gamma(h_a\rightarrow X_mX_n)$. Heavier modes are even broader due to
the increase in available final
states. Similar to the hidden vector decays the presence of an
approximate KK 
number conservation means that decays with $a\sim m+n$ are dominant,
as can be seen  in
Fig.~\ref{fig:graviton_a_45_width_v_daughterkk_number_nohiggs} where
we plot $\Gamma (h_{a=45}\rightarrow X_mX_n)$ against daughter KK
number for the Higgsless case with fixed $m+n$.
%--------------------------------------------------------%
%-----------------------------------------------------------%

Note that crossing symmetry requires the matrix element for the KK graviton decay and the KK vector decay to be related and one can show that this requires the amplitudes for decay to be related via
\begin{eqnarray}
\left.\sum_{pol.}|\mathcal{M}(X_n \rightarrow h_aX_m)|^2\right|_{\xi\rightarrow-\xi}=\sum_{pol.}|\mathcal{M}(h_a \rightarrow X_nX_m)|^2.
\end{eqnarray}
We have checked that this relation holds for the matrix elements we
have employed. We also note that, in general, stabilization of the extra dimension will involve some
additional bulk field like a Goldberger-Wise
scalar~\cite{Goldberger:1999uk} or some
flux. This will couple to KK gravitons and one expects additional
decay channels as a result. 

%%%%%%%%%%%%%%%%%%%%%%%%%%%%%%%%%%55

\section{Amplitude for $e^+e^-\to f\bar{f}$\label{app:pew}}
Using the vertices from Sec.~\ref{pew_e-->f} we can calculate the
process $e^+e^-\to
f\bar{f}$ with $f\neq e$. We work to leading non-trivial
order in $\epsilon_*$ and neglect the
fermion masses.  The resulting summed and squared matrix element is: 
\bea
\frac{1}{4}\sum_{s,s'}|\mathcal{M}|^2 &=& 
\lrf{s^2}{4~}\,\left[
\left(|a_{LL}|^2+|a_{RR}|^2+|a_{LR}|^2+|a_{RL}|^2\right)(1+\cos^2\theta)\right.
\\
&&
~+~ \left.\left(|a_{LL}|^2+|a_{RR}|^2-|a_{LR}|^2-|a_{RL}|^2\right)\,
(2\,\cos\theta)
\right],
\nnmb
\eea
where the factors $a_{AB}$ ($A\,B\!=\!L,R$) are given by
\beq
a_{AB} = (a_Z+a_{ZXZ})\,g_{A_Z}^eg_{B_Z}^f 
+ (a_{\gamma}+a_{\gamma X\gamma})\,g_{A_{\gamma}}^eg_{B_{\gamma}}^f
+a_{ZX\gamma}\,g_{A_{Z}}^eg_{B_{\gamma}}^f
+a_{\gamma XZ}\,g_{A_{\gamma}}^eg_{B_{Z}}^f.
\eeq
Here, $g^f_{L_{\gamma}} = g^f_{R_{\gamma}} = e\,Q= (\bar{g}/c_Ws_W)\,Q$,
$g^f_{L_Z} = \bar{g}(t^3_L-Q\,s_W^2)$, and $ g^f_{R_Z} = \bar{g}(-Q\,s_W^2)$,
$\bar{g} = \sqrt{g^2+g'^2}$,
while the $a_{V}$ terms are given by
\bea
a_Z &=& 1/(p^2-m_z^2+i\Gamma_Zm_Z) := \Delta_p^Z\\
a_{ZXZ} &=& \frac{\epsilon_*^2}{M_*}s_W^2p^4(\Delta_p^Z)^2\Delta_p^{UV}\\
a_{\gamma} &=& \frac{1}{p^2}\\
a_{\gamma X\gamma} &=& \frac{\epsilon_*^2}{M_*}c_W^2\Delta_p^{UV}\\
a_{ZX\gamma} &=& \frac{\epsilon_*^2}{M_*}(-c_Ws_W)\,{p^2}\Delta_p^{Z}
\Delta_p^{UV} 
= a_{\gamma XZ}.
\eea
Here, $\Delta_p^{UV}$ is the 5D bulk-to-bulk propagator for which we
use the matched expression of Eq.~\eqref{prop_emp}.  When the final
state is $e^+e^-$, there is an additional $t$-channel contribution
to the amplitude.
%--------------------------------------------------------%
%-----------------------------------------------------------%
%%%%%%%%%%%%%%%%%%%%%%%%%%%%%%%%%%%%%%%%%%%%%%%%%%%%%%%%%%%%%%%%%

%\newpage

\end{document}